\documentclass[amssymb,amsmath,superscriptaddress,footinbib,twocolumn]{revtex4}
\usepackage{natbib,graphicx,subfigure,subeqnarray,fancyhdr,amsmath,multirow,color, mathrsfs}
\begin{document}

\renewcommand{\labelitemi}{$-$}
\newcommand{\change}[1]{{\color{black} #1}}
\newcommand{\Fc}{\mathcal{F}}\newcommand{\Rc}{\mathcal{R}}\newcommand{\dd}{\mathrm{d}}
\newcommand{\ee}{\mathrm{e}}\newcommand{\ci}{\mathrm{i}}\newcommand{\ib}{\mathbf{i}}
\newcommand{\jb}{\mathbf{j}}\newcommand{\kb}{\mathbf{k}}\newcommand{\ab}{\mathbf{a}}
\newcommand{\Fb}{\mathbf{F}}\newcommand{\fb}{\mathbf{f}}\newcommand{\Gb}{\mathbf{G}}
\newcommand{\Mb}{\mathbf{M}Ä}\newcommand{\nb}{\mathbf{n}}\newcommand{\Sb}{\mathbf{S}}
\newcommand{\Sbs}{\mathbf{S^*}}\newcommand{\Rb}{\mathbf{R}}\newcommand{\Sigb}{\boldsymbol{\Sigma}}\newcommand{\sigb}{\boldsymbol{\sigma}}
\newcommand{\Sigbs}{\boldsymbol{\Sigma^*}}\newcommand{\alphab}{\boldsymbol\alpha}
\newcommand{\omegab}{\boldsymbol{\omega}}
\newcommand{\epsb}{\boldsymbol{\epsilon}}
\newcommand{\ub}{\mathbf{u}}
\newcommand{\xib}{\boldsymbol{\xi}}\newcommand{\eb}{\mathbf{e}}\newcommand{\vv}[1]{\underline{#1}}\newcommand{\ev}{\vv{e}}
\newcommand{\rv}{\vv{r}}\newcommand{\TT}[1]{\underline{\underline{#1}}}\newcommand{\omb}{\mathbf{\omega}}
\def\v{\vspace{3cm}}
\newcommand{\Ub}{\mathbf{U}}\newcommand{\xb}{\mathbf{x}}\newcommand{\rb}{\mathbf{r}}
\newcommand{\ssb}{\mathbf{s}}\newcommand{\Xb}{\mathbf{X}}\newcommand{\Pe}{\mbox{Pe}}\newcommand{\Da}{\mbox{Da}\,}
\newcommand{\mean}[1]{\langle #1\rangle}
\newcommand{\ddp}{[p]^\pm}\newcommand{\taub}{\mbox{\boldmath$\tau$}}\newcommand{\Fr}{\mbox{\textit{Fr}}}
\let\grad\nabla\newcommand{\z}{\zeta}\newcommand{\kk}{\kappa}\newcommand{\tkk}{\tilde{\kappa}}
\newcommand{\e}{\varepsilon}\newcommand{\zb}{\bar{\zeta}}\let\grad\nabla\let\bcdot\cdot
\newcommand{\half}{{\textstyle\frac{1}{2}}}
\newcommand{\textfrac}[2]{{\textstyle\frac{#1}{#2}}}
\newcommand{\LF}[1]{{#1}^{\mathrm{LF}}}\newcommand{\Lap}[1]{{#1}^{\mathrm{L}}}
\newcommand{\ds}{*\!*}\newcommand{\cond}[2]{\frac{\mathrm{D} #1}{\mathrm{D} #2}}
\newcommand{\pard}[2]{\frac{\partial #1}{\partial #2}}\newcommand{\totd}[2]{\frac{\mathrm{d}#1}{\mathrm{d}#2}}
\newcommand{\pardd}[3]{\frac{\partial^2 #1}{\partial #2 \partial #3}}
\newcommand{\Rey}{\mbox{Re}}\newcommand{\Imag}{\mbox{Im}}
\newcommand{\Fpint}{=\!\!\!\!\!\!\!\int}
\newcommand{\txi}{\tilde\xi}\newcommand{\dxi}{\delta\xi}
\newcommand{\tpsi}{\tilde\psi}\newcommand{\dpsi}{\delta\psi}
\makeatletter
\def\sgn{\mathop{\operator@font sgn}}
\makeatother
\allowdisplaybreaks[1]

\title{Autophoretic locomotion from geometric asymmetry}
\author{S\'ebastien Michelin}
\email{sebastien.michelin@ladhyx.polytechnique.fr}
\affiliation{LadHyX -- D\'epartement de M\'ecanique, Ecole Polytechnique -- CNRS, 91128 Palaiseau, France}
\author{Eric Lauga}
\email{e.lauga@damtp.cam.ac.uk}
\affiliation{Department of Applied Mathematics and Theoretical Physics, University of Cambridge, Cambridge CB3 0WA, United Kingdom}
\date{\today}

\begin{abstract}
Among the few methods which have  been proposed to create small-scale swimmers, those relying on self-phoretic mechanisms present an interesting design challenge in that  chemical gradients are required  to generate net propulsion. Building on recent work, we propose that asymmetries in geometry are sufficient to induce chemical gradients and swimming. We illustrate this idea using two different calculations. We first calculate exactly the self-propulsion speed of a system composed of two spheres of unequal sizes but identically chemically homogeneous. We then consider arbitrary, small-amplitude, shape deformations of a chemically-homogeneous sphere, and calculate asymptotically  the self-propulsion velocity induced by the shape asymmetries. Our results demonstrate how geometric asymmetries can be tuned to induce large locomotion speeds without the need of chemical patterning. 
 \end{abstract} 
\maketitle
%%%%%%%%%%%%%%%%%%%%%%
\section{Introduction}
\subsection{Background}

Achieving self-propulsion at the micro-scale is essential to many biological organisms and  functions, including   migration, feeding or escaping aggressions \citep{stocker2008,hamel2011}, and  reproductive success of other larger species (e.g.~mammals \citep{suarez2006}).   From an engineering point of view, the design of self-propelled systems or micro-/nano-robots may offer important opportunities in particular for biomedical applications, to perform controlled therapeutic or diagnostic tasks \cite{Wang2009,Nelson2010}.

Many efforts dedicated to the design of such microscale artificial ``swimmers'' have been inspired by the biological world, where  viscous locomotion is achieved in the absence of inertial forces  \citep{lauga2009}, for example using rotation or waving of rigid or flexible filaments. 
Experimentally, three broad categories of synthetic swimmers have been realized so far: (i) rigid \cite{ghosh2009,Zhang2010b,tottori12} or flexible  helices  \cite{dreyfus05,Gao2010,Pak2011} or planar filaments \cite{dreyfus05}, inspired by bacterial flagella, forced by an external  magnetic field in order to achieve propulsion (see also Ref.~\citep{tierno10} for a variation  using nearby surfaces); (ii) rigid bodies moving under the action of an external standing-wave acoustic field \cite{wang2012,nadal2014};  and (iii) so-called phoretic swimmers \cite{paxton2004}. 
The first two critically rely on the existence of an outside forcing in order to move,  which may not only limit their applications but also disqualify them from achieving force- and torque-free propulsion. In contrast,   
phoretic (or fuel-based) locomotion, which is the focus of the present study,  relies solely on the interaction of a rigid body with the solute content in the    surrounding fluid.

The ability to generate an effective slip velocity along a solid boundary outside a thin interaction layer in response to a local tangential solute  gradient is at the heart of classical phoretic physics. It  originates from local pressure imbalance which are a consequence of short-range  solute-particle  interactions \citep{anderson1989}. Classically, this phoretic mobility is responsible for the migration of inert particles in externally-applied chemical gradients. Specifically, a particle with uniform local surface mobility ${\cal M}$ placed in a far-field chemical gradient ${\bf G}$ \change{ of a neutral solute} experiences a distribution of slip velocity on its surface leading to a global phoretic velocity $\Ub=-{\cal M}{\bf G}$. \change{For other phoretic mechanisms such as diffusiophoresis of charged solutes or thermophoresis, the slip and phoretic velocities  depend instead  on the gradient of the logarithm of the concentration/temperature, and this linear relationship is only observed for sufficiently small gradients \cite{anderson1989}.}  The basic idea of autophoresis is to combine such  phoretic mobility with a chemical surface activity. Using chemical reactions catalyzed at its surface, a chemically-active particle  becomes able to generate the tangential gradients necessary to achieve its own propulsion \citep{golestanian2005}.

The feasibility of  self-diffusiophoresis was recently demonstrated in several experimental studies using the catalytic decomposition of hydrogen peroxide on platinum-coated surfaces \citep{Howse2007,Ebbens2011,ebbens2012}, although the exact physico-chemical mechanism at play is still under debate \citep{brown2014,ebbens2014}. Note that this mechanism
%, and the coupling of a solute transport to the phoretic flows around the particle through the activity and mobility surface properties, 
shares several important similarities with self-thermophoresis~\citep{jiang2010} or the self-propulsion of droplets through Marangoni effects, for which the slip velocity is replaced with a surface shear stress discontinuity \citep{thutupalli2011,schmitt2013,izri2014}.

In order to generate self-propulsion,  breaking  spatial symmetries is required. Indeed,  the diffusion of a  solute outside a homogeneous spherical particle leads to a spherically-symmetric  concentration, and thus no slip velocity and no self-propulsion. Two routes have been identified to break  symmetries and enable propulsion: (i) the chemical patterning  on the surface of the particle, as used in most experiments with ``Janus'' particles \cite{Howse2007,Ebbens2011,theurkauff2012,ebbens2012}; (ii) an instability mechanism resulting from the nonlinear advective coupling of the solute to the phoretic flows it creates, which  spontaneously break symmetries and propel isotropic particles or droplets \cite{michelin2013c,izri2014}.

An alternative route to symmetry-breaking originates solely from geometry. Consider a particle with homogeneous surface properties (i.e.~uniform surface activity and mobility). In the absence of inertia, an asymmetry in the particle  shape will in general create non-homogeneous solute (or reactant) concentrations along its boundary, which are then  likely to have non-zero average and therefore lead to propulsion.  This idea is at the heart of  experiments on collective phoretic dynamics \cite{palacci2013} and it was recently  the focus of an  article   analyzing the self-propulsion of a near-sphere with low-order azimuthal perturbations \citep{shklyaev2014} using the osmotic framework of Brady and coworkers \cite{cordova2008,brady2011}.
In this paper, we  tackle the same problem within the classical continuum framework of self-diffusiophoresis \cite{golestanian2007,julicher2009,sabass2012}, focusing on two classical geometries amenable to analytical calculations.

\subsection{Intuitive model}

Before presenting detailed calculations in the following sections, we illustrate here intuitively the idea of acquiring locomotion from shape asymmetries by considering the case of a swimmer composed of two rigid spheres which share the same uniform chemical surface properties. The spheres have radii $R_1$ and $R_2$, and their centers are separated by a distance $d$. We are going to show that phoretic  locomotion is guaranteed provided $R_1\neq R_2$. 

\begin{figure*}
\begin{center}
\begin{tabular}{cc}
\includegraphics[height=7cm]{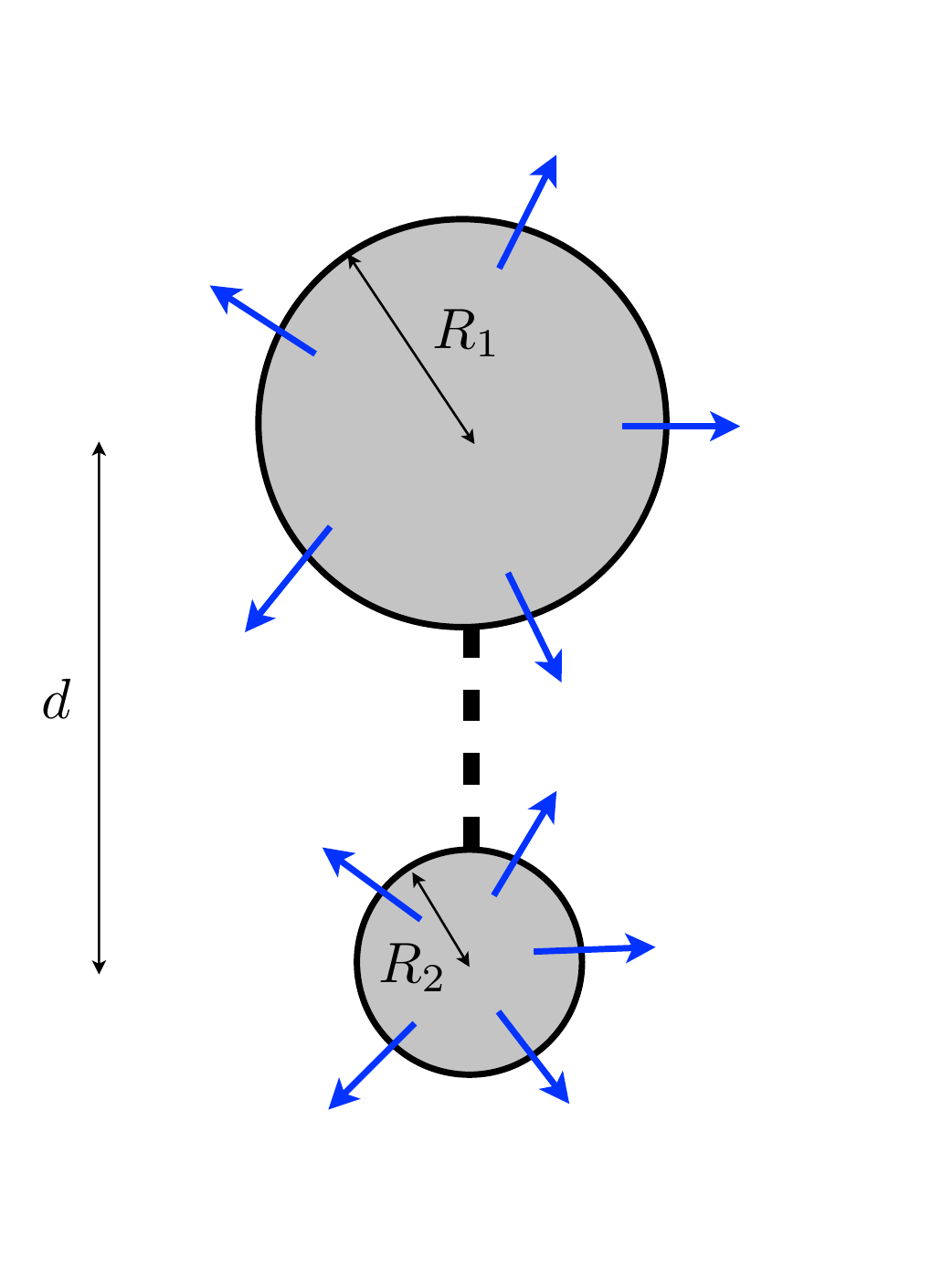} &\hspace{1cm} \includegraphics[height=7cm]{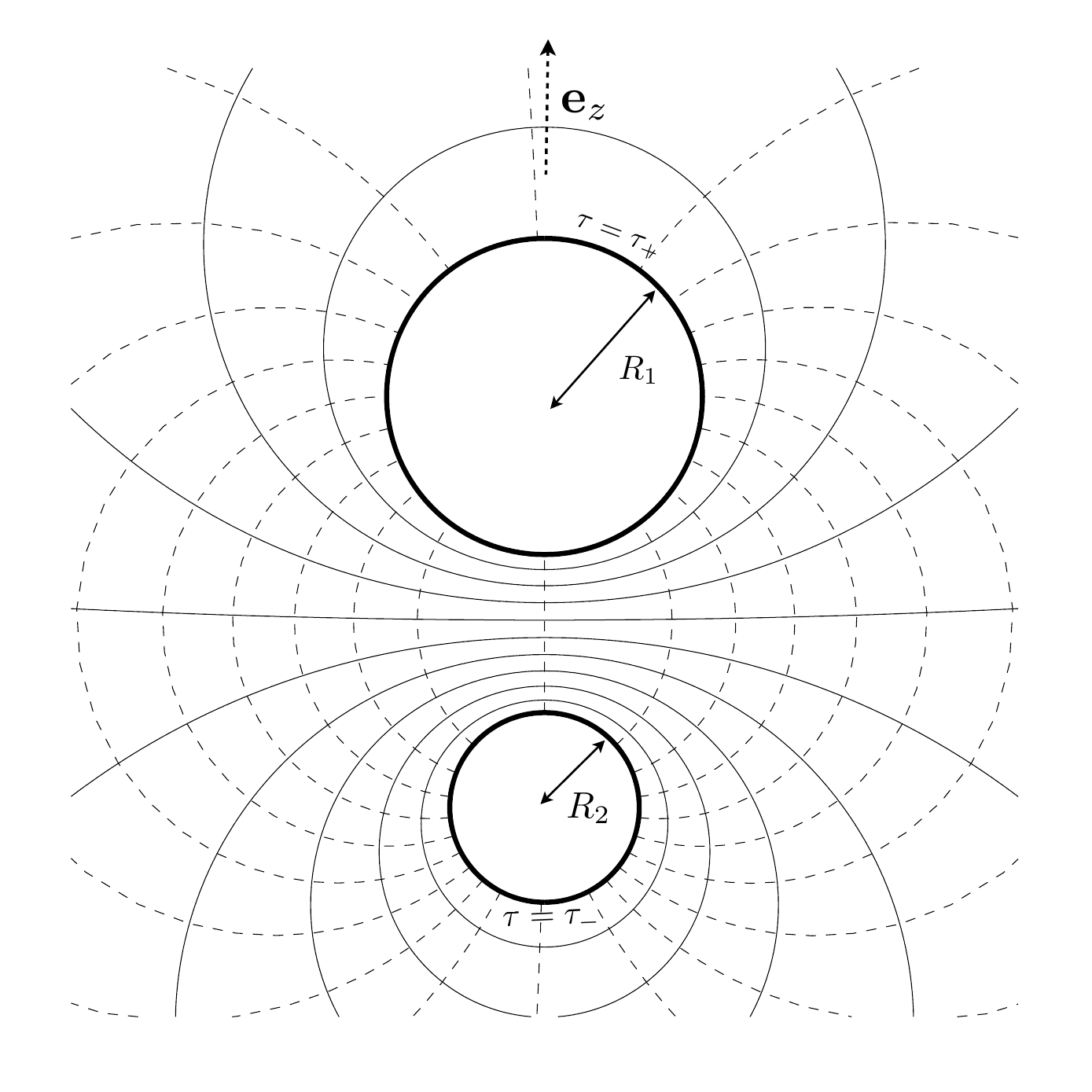}
\end{tabular}
\caption{Left: Notation for a two-sphere auto-phoretic system. The surface activity is uniform on both spheres, corresponding to a rate of solute release/absorption equal to $\mathcal{A}$. Right: Bi-spherical coordinates: surfaces of constant $\tau$ (solid) and $\mu$ (dashed) are shown; the surfaces $\tau=\tau_\pm$ correspond to the boundaries of the spheres. }\label{fig:schema}
\end{center}
\end{figure*}

Each sphere emits a solute with a uniform and identical rate $\cal A$ which diffuses in the fluid domain with diffusion constant $\kappa$. We denoted by $\eb_z$ the unit vector joining the spheres' centers (Fig.~\ref{fig:schema}). 
In the limit of large separation between the spheres,  $d\gg R_i$, the leading order concentration field can be obtained by superposition of the distribution generated by each sphere independently
\begin{equation}
c(\rb)=\frac{\cal A}{\kappa}\left(\frac{R_1^2}{|\rb-\rb_1|}+\frac{R_2^2}{|\rb-\rb_2|}\right),
\end{equation}
with corrections arising at higher order in $1/d$. Each sphere  is then exposed to two different phoretic  contributions: firstly its own concentration field  but since this  is isotropic, it does not lead to any surface gradient or slip velocity; secondly the concentration gradients generated by the other sphere. Assuming a constant phoretic mobility $\cal M$ relating flow velocities to chemical gradients,   the  propulsion velocities of each spheres, $\Ub^{f}_1$ and $\Ub^{f}_2$, which would arise if they were individually force-free are given by
\begin{equation}
\Ub_1^{f}=\frac{{\cal A}{\cal M}R_2^2}{\kappa d^2}\eb_z,\qquad \Ub_2^{f}=-\frac{{\cal A}{\cal M}R_1^2}{\kappa d^2}\eb_z.
\end{equation}
A rigid two-sphere system (i.e. where $d$ is kept constant) moves thus at speed $\Ub$ in a Newtonian fluid of viscosity $\eta$ such that the total hydrodynamic force is zero. In the far-field limit, the hydrodynamic resistance of sphere $i$ is $6\pi\mu R_i$ and the total hydrodynamic force acting on the fluid is $6\pi\eta R_2 (\Ub-\Ub_2^f)+6\pi\eta R_1(\Ub-\Ub_1^f)=\bf 0$
leading to locomotion at speed
\begin{equation}\label{eq:farfieldresult}
\Ub=\frac{{\cal A}{\cal M} R_1R_2(R_2-R_1)}{\kappa d^2(R_2+R_1)}\eb_z.
\end{equation}
As long as $R_1\neq R_2$,  a net phoretic velocity is therefore induced. The origin of this velocity is purely in the chemical asymmetries resulting from geometric differences, since both spheres are similarly  chemically homogeneous. The purpose of this paper is to illustrate and analyze these ideas further using two exact calculations.

\subsection{Outline of the paper}

In this paper, we study in detail how  shape asymmetries can lead to  autophoretic locomotion. In order to combine physics with more practical considerations, we focus on two specific geometries. The first one, studied in detail in  \S\ref{sec:two_spheres}, determines the exact solution for the homogenous two-sphere system introduced above. This geometry allows us to address large shape asymmetries and can provide the basis for further experimental investigations.  
The second geometry is that of a near sphere, addressed in  
\S\ref{sec:near_sphere}, which allows us to quantify which asymmetric  surface modes play a role in the transition to locomotion and to address the question of optimal self-propulsion.

\subsection{Phoretic continuum framework and scaling}

Both problems in this paper  are treated within the continuum framework of autophoretic propulsion \citep{golestanian2007,julicher2009,sabass2012}, by considering an isolated rigid system $\cal S$ (a two-sphere system in \S\ref{sec:two_spheres}  or a near-sphere in \S\ref{sec:near_sphere}) in an unbounded fluid domain of dynamic viscosity $\eta$ and density $\rho$. The rigid body interacts  with a solute species of local concentration $C$ that diffuses in the fluid medium with diffusivity $\kappa$. \change{The interaction layer thickness $\lambda$ is assumed to be small enough for the classical slip-velocity formulation to be valid \cite{michelin2014}.} The particle's surface chemical properties are \change{here} characterized by a homogeneous surface   activity $\cal A$ and mobility $\cal M$. At the surface of the particle, the solute is thus released (${\cal A}>0$) or absorbed (${\cal A}<0$) with a fixed flux, so that
\begin{equation}
\kappa\nb\cdot\grad c=-{\cal A}\quad \textrm{on  }\cal S,
\end{equation}
and a slip velocity arises proportional to the local solute concentration gradient along the surface, following
\begin{equation}
\ub={\cal M}(1-\nb\nb)\cdot\grad c\quad \textrm{on   }\cal S.\label{eq:bc_phoretic}
\end{equation}

If advection of the solute by the phoretic flows can be neglected (namely if the P\'eclet number, $\Pe=\cal{U}R/\kappa$, is sufficiently small, with $\cal{U}$ the typical velocity magnitude and $R$ the characteristic size of the solid system), the solute has a purely diffusive behavior outside $\cal S$, so that the concentration relative  to the far-field solute content $c=C-C_\infty$ satisfies
\begin{equation}
\nabla^2 c=0.
\end{equation}
Because the typical particle size and  phoretic flow magnitude are small, inertial effects in the flow dynamics can be neglected, i.e.~the Reynolds number $\Rey=\rho \cal{U}R/\eta$ is small. The flow field resulting from the phoretic slip at the boundary can then be solved for, in the reference frame attached to the particle, using Stokes' equations
\begin{equation}
\eta\nabla^2\ub=\grad p,\quad \grad\cdot\ub=0,
\end{equation}
together with the mobility condition in Eq.~\eqref{eq:bc_phoretic}. At infinity,  we have 
\begin{equation}
\ub(\rb\rightarrow\infty)\sim-(\Ub+\boldsymbol\Omega\times\xb),
\end{equation}
 where $(\Ub,\boldsymbol\Omega)$ are the self-propulsion velocity and rotation of the solid system. As both the solid and fluid inertia are negligible in the Stokes limit,  the solid system must remain force-free, a condition that uniquely determines the self-propulsion kinematics $(\Ub,\boldsymbol\Omega)$. Since we focus in the following on axisymmetric systems, we have by symmetry $\boldsymbol\Omega=\bf 0$ and $\Ub$ is parallel to the axis of symmetry.

The problem is non-dimensionalized using $R$ as characteristic length scale (the radius of one sphere in the two-sphere system or the mean radius of the near-sphere), $|\cal AM|/\kappa$ as characteristic velocity, $|\cal A| R/\kappa$ as characteristic concentration fluctuation and $\eta|\cal AM|/R\kappa$ as characteristic pressure. In the purely diffusive limit ($\Pe=0$), the only non-dimensional parameters arise from geometry. 

This fixed-flux approach can be generalized to a simple one-step chemical reaction where the solute is consumed (${\cal A}<0$) at the surface at a rate proportional to its concentration \citep{michelin2014}. In that case, ${\cal A}=-{\cal K} c$ where the reaction rate $\cal K$ is now the uniform chemical property. The typical activity is now ${\cal K} C_\infty$. This introduces an additional non-dimensional parameter, the Damk\"ohler number $\Da={\cal K R}/\kappa$, which quantifies the limitation of the reaction rate by diffusion. For $\Da\ll 1$, diffusion is sufficiently fast to replenish the solute content in the vicinity of the sphere and the solute consumption occurs approximately at a fixed rate, while for $\Da\gg 1$, the reaction is limited by the depletion of the solute content which cannot be compensated by  the slow diffusion.

%%%%%%%%%%%%%%%%%%%%%%
\section{Autophoretic locomotion of  a homogeneous  two-sphere system}
\label{sec:two_spheres}

\subsection{Problem formulation}
We consider a system consisting of two spheres, $\mathcal{S}_1$ and $\mathcal{S}_2$, of respective radii $R_1$ and $R_2$, with a fixed distance $d$ maintained between their respective centers (either through long-range interactions or through a connecting rod with negligible hydrodynamic influence). \change{In contrast with existing studies focusing on a geometrically-symmetric dimer with chemical asymmetry \cite{ruckner2007,popescu2011}, we specifically investigate here the effect of the geometric asymmetry (i.e.~different radii) for a chemically-symmetric system.} Both spheres have uniform surface properties (activity and mobility)  so that  the axis joining their centers $\eb_z$ is an axis of symmetry for the problem. We first  seek a solution of the diffusion problem for the solute concentration (relative to its far-field value)
\begin{align}
&\nabla^2c=0\textrm{      outside the spheres},\\
&\nb\cdot\nabla c=-A \textrm{      on   } \mathcal{S}_1 \textrm{  and  }\mathcal{S}_2,\\
&c(r\rightarrow\infty)\rightarrow 0,
\end{align}
with $A={\cal A}/|{\cal A}|=\pm 1$, the dimensionless activity. Concentration gradients at the surface of the two spheres generate surface slip velocities, 
and the following hydrodynamics problem must then be solved
\begin{align}
&\nabla^2\ub=\nabla p\textrm{      outside the spheres},\\
&\ub=\ub_s=M(\mathbf{1}-\nb\nb)\cdot\grad c \textrm{      on   } \mathcal{S}_1 \textrm{  and  }\mathcal{S}_2,\\
&\ub(r\rightarrow\infty)\sim -U\eb_z,
\end{align}
where the swimming velocity $U$ is such that the total hydrodynamic force on the two-sphere system is zero  and $M={\cal M}/|{\cal M}|=\pm 1$ is the dimensionless  mobility.

Alternatively, using the reciprocal theorem for Stokes flows \cite{stone1996}, the swimming velocity can be determined directly from the slip velocity distribution $\ub_s$ on the two spheres as
\begin{equation}
U=-\frac{1}{F^*}\int_{\mathcal{S}_1,\mathcal{S}_2}\ub_s\cdot\boldsymbol\sigma^*\cdot\nb\,\dd S,\label{eq:reciprocal}
\end{equation}
where $(\ub^*,\sigb^*)$ is the solution of the dual hydrodynamic problem obtained by imposing a steady velocity $U^*\eb_z$ to the equivalent rigid (no-slip) two-sphere system
\begin{align}
&\nabla^2\ub^*=\nabla p^*\textrm{      outside the spheres},\\
&\ub=U^*\eb_z \textrm{      on   } \mathcal{S}_1 \textrm{  and  }\mathcal{S}_2,\\
&\ub^*(r\rightarrow\infty)\rightarrow 0,
\end{align}
 corresponding thus to a total hydrodynamic force $F^*$ 
\begin{align}
&F^*=\int_{\mathcal{S}_1,\mathcal{S}_2}\eb_z\cdot\sigb^*\cdot\nb\,\dd S,
\end{align}
with  $\sigb^*=-p^*\mathbf{1}+\left(\grad \ub^*+\grad\ub^{*T}\right)$.

\subsection{Computing the swimming velocity}
\subsubsection{Bi-spherical geometry}
Taking advantage of the  geometry  of the problem and its symmetries, a bi-spherical polar coordinate system $(\tau,\mu,\phi)$ is used. Noting $(\rho,\phi,z)$ the classical cylindrical polar coordinates, the bi-spherical coordinates $(-\infty<\tau<\infty,-1\leq\mu\leq 1)$ are defined as
\begin{equation}
\rho=\frac{a\sqrt{1-\mu^2}}{\cosh\tau-\mu},\quad z=\frac{a\sinh\tau}{\cosh\tau-\mu}\cdot
\end{equation}
In this system of coordinates, $\tau=\tau_0$ is a sphere centered on the $z$-axis at $z=a\coth\tau_0$ with  radius $a/|\sinh\tau_0|$. The origin of the system of coordinates as well as the constant $a>0$ are chosen such that $\tau=\tau_+>0$ (resp. $\tau=\tau_-<0$) corresponds to the surface of $\mathcal{S}_1$ (resp. $\mathcal{S}_2$) located in the upper (resp. lower) half-plane. The physical parameters $(d,R_1,R_2)$ can be expressed in terms of $(a,\tau_-,\tau_+)$ as
\begin{align}
R_2=-\frac{a}{\sinh\tau_-},\, R_1=\frac{a}{\sinh\tau_+},\, d=a (\coth\tau_+-\coth\tau_-),
\end{align}
and provided that $d\geq R_1+R_2$, the solution of the inverse system is unique. The unit vectors $\eb_\tau$ and $\eb_\mu$, respectively normal to surfaces of constant $\tau$ and $\mu$, are defined by $\partial\xb/\partial\tau=h_\tau\eb_\tau$ and $\partial\xb/\partial \mu=h_\mu\eb_\mu$ with $h_\tau$ and $h_\mu$ the metric coefficients
\begin{equation}\label{eq:metric}
h_\tau=\frac{a}{\cosh{\tau}-\mu}, \quad h_\mu=\frac{a}{(\cosh\tau-\mu)\sqrt{1-\mu^2}}\cdot
\end{equation}
Consequently we have 
\begin{align}
\eb_\tau=\frac{1-\mu\cosh\tau}{\cosh\tau-\mu}\eb_z-\frac{\sqrt{1-\mu^2}\sinh\tau}{\cosh\tau-\mu}\eb_\rho,\\
\eb_\mu=\frac{\sqrt{1-\mu^2}\sinh\tau}{\cosh\tau-\mu}\eb_z+\frac{1-\mu\cosh\tau}{\cosh\tau-\mu}\eb_\rho,
\end{align}
and $(\eb_\tau,\eb_\mu,\eb_\phi)$ form a direct orthonormal basis.

\subsubsection{Solving the solute diffusion equation}
Within this system of coordinates, the Laplace equation for the concentration field becomes
\begin{equation}
\pard{}{\tau}\left(\frac{1}{\cosh\tau-\mu}\pard{c}{\tau}\right)+\pard{}{\mu}\left(\frac{1-\mu^2}{\cosh\tau-\mu}\pard{c}{\mu}\right)=0,
\end{equation}
whose general solution  vanishing at infinity is \cite{stimson1926}
\begin{equation}\label{eq:cdef}
c(\tau,\mu)=\sqrt{\cosh\tau-\mu}\sum_{n=0}^\infty c_n(\tau)L_n(\mu),
\end{equation}
with $L_n$ the Legendre polynomial of degree $n$, and 
\begin{eqnarray}\label{eq:csol}
\notag c_n(\tau)&=&a_n\mathrm{exp}\left[\left(n+\frac{1}{2}\right)(\tau-\tau_+)\right]\\
&& +b_n\mathrm{exp}\left[-\left(n+\frac{1}{2}\right)(\tau-\tau_-)\right].
\end{eqnarray}
The constants, $a_n$ and $b_n$ are determined by imposing the flux boundary condition on each sphere (noting that for $\tau=\tau_\pm$, $\nb(\tau_\pm)=\mp\eb_\tau$)
\begin{equation}
\eb_\tau\cdot\grad c=\frac{1}{h_\tau}\pard{c}{\tau}=\pm A \textrm{   at   }\tau=\tau_\pm.
\end{equation}

Substituting Eqs.~\eqref{eq:metric} and \eqref{eq:cdef}, and  projecting the previous equation along the Legendre polynomial $L_p(\mu)$, and using formulae summarized in Appendix~\ref{sec:app_formulas} we obtain 
\begin{eqnarray}\notag 
\cosh\tau_\pm c_p'(\tau_\pm)-\frac{p}{2p-1}c_{p-1}'(\tau_\pm)-\frac{p+1}{2p+3}c_{p+1}'(\tau_\pm)\\ 
 +\frac{\sinh\tau_\pm}{2}c_p(\tau_\pm)=\pm aA\sqrt{2}\ee^{-(p+1/2)|\tau_\pm|}.\quad
\end{eqnarray}
Together with Eq.~\eqref{eq:csol}, the previous equation form a linear system for the coefficients $a_n$, $b_n$ which can be solved numerically after truncating the sum in Eq.~\eqref{eq:cdef}.

Once these constants have been determined, the slip velocity $\ub_s$ can be computed on the surface of each sphere as
\begin{align}
u_\mu^s(\tau_\pm,\mu)&=\frac{M}{a}\sqrt{1-\mu^2}(\cosh\tau_\pm-\mu)\pard{c}{\mu}\nonumber\\
&=\frac{M\sqrt{1-\mu^2}}{a}\sum_{n=0}^\infty c_n(\tau_\pm)\notag\times\\
&\left[(\cosh\tau_\pm-\mu)^{3/2}L_n'(\mu)-\frac{L_n(\mu)}{2}\sqrt{\cosh\tau_\pm-\mu}\right].
\end{align}

\subsubsection{Solving the dual rigid-body problem}
The solution of the general Stokes flow problem for an axisymmetric problem in bi-spherical coordinates can be expressed in terms of the streamfunction $\psi(\tau,\mu)$, 
\begin{equation}
u_\tau=(\cosh\tau-\mu)^2\pard{\psi}{\mu},\qquad u_\mu=-\frac{(\cosh\tau-\mu)^2}{\sqrt{1-\mu^2}}\pard{\psi}{\tau}.
\end{equation}
The solution of the dual problem required to solve Eq.~\eqref{eq:reciprocal}, i.e.~Stokes flow vanishing at infinity with imposed velocity $U^*\eb_z$ on the spheres, is given by \cite{stimson1926}
\begin{equation}\label{eq:psi}
\frac{\psi^*(\tau,\mu)}{U^*}=(\cosh\tau-\mu)^{-3/2}\sum_{n=1}^\infty(1-\mu^2)L_n'(\mu)U_n(\tau),
\end{equation}
with
\begin{eqnarray}
U_n(\tau)&=&\alpha_n\cosh\left(n+\frac{3}{2}\right)\tau +\beta_n\sinh\left(n+\frac{3}{2}\right)\tau\notag\\&&
+\gamma_n\cosh\left(n-\frac{1}{2}\right)\tau+\delta_n\sinh\left(n-\frac{1}{2}\right)\tau.\quad
\end{eqnarray}
In the previous equation, the four sets of constants $\alpha_n$, $\beta_n$, $\gamma_n$ and $\delta_n$ are determined by imposing the no-slip boundary condition on both spheres $\ub^*=U^*\eb_z$ in the dual problem or equivalently
\begin{subeqnarray}\label{eq:bcpsi}
\pard{\psi^*}{\mu}(\tau_\pm,\mu)=\frac{U^*(1-\mu\cosh\tau_\pm)}{(\cosh\tau_\pm-\mu)^3},\\ \pard{\psi^*}{\tau}(\tau_\pm,\mu)=-\frac{U^*(1-\mu^2)\sinh\tau_\pm}{(\cosh\tau_\pm-\mu)^3}.
\end{subeqnarray}
These equations can be rewritten using Eq.~\eqref{eq:psi} as
\begin{subeqnarray}
&&\frac{(1-\mu\cosh\tau_\pm)}{\sqrt{\cosh\tau_\pm-\mu}} = \sum_{n=1}^\infty U_n(\tau_\pm)\times\\
&&\notag\left[-n(n+1)L_n(\mu)(\cosh\tau_\pm-\mu)+\frac{3}{2}(1-\mu^2)L_n'(\mu)\right],\\
&&-\frac{(1-\mu^2)\sinh\tau_\pm}{\sqrt{\cosh\tau_\pm-\mu}} = \sum_{n=1}^\infty (1-\mu^2)L_n'(\mu)\times\\
&&\notag\left[U_n'(\tau_\pm)(\cosh\tau_\pm-\mu)-\frac{3\sinh\tau_\pm}{2}U_n(\tau_\pm)\right].
\end{subeqnarray}

Projecting the two previous equations onto $L_p(\mu)$ and $L_p'(\mu)$ respectively, and using the relations given in Appendix~\ref{sec:app_formulas}, we finally obtain
\begin{align}
 \nonumber -p(p+1)\cosh\tau_\pm U_p(\tau_\pm)+  \nonumber \frac{p(p-1)(2p-3)}{2(2p-1)}&U_{p-1}(\tau_\pm)\\
 +\frac{(p+1)(p+2)(2p+5)}{2(2p+3)}U_{p+1}(\tau_\pm)&=f_p(\mu),\label{eq:un}\\
\cosh\tau_\pm U_p'(\tau_\pm)-\frac{p-1}{2p-1}U_{p-1}'(\tau_\pm)-\frac{p+2}{2p+3}&U_{p+1}'(\tau_\pm)\nonumber \\
-\frac{3}{2}\sinh\tau_\pm U_p(\tau_\pm)&=\tilde{f}_p(\mu)\label{eq:dun},
\end{align}
with 
\begin{align}
f_p(\mu)&=\sqrt{2}\left(\ee^{-(p+1/2)|\tau_\pm|}-\frac{(p+1)\cosh\tau_\pm}{2p+3}\ee^{-(p+3/2)|\tau_\pm|}\right.\nonumber\\
&\left.-\frac{p\cosh\tau_\pm}{2p-1}\ee^{-(p-1/2)|\tau_\pm|}\right),\\
\tilde{f}_p(\mu)&=\sqrt{2}\sinh(\tau_\pm)\left[\frac{\ee^{-(p+3/2)|\tau_\pm|}}{2p+3}-\frac{\ee^{-(p-1/2)|\tau_\pm|}}{2p-1}\right].
\end{align}

Using Eqs.~\eqref{eq:psi}, \eqref{eq:un} and \eqref{eq:dun} applied at $\tau_\pm$ provides independent sets of four linear equations for the integration constants $\alpha_n$, $\beta_n$, $\gamma_n$ and $\delta_n$, which can be obtained numerically.

\subsubsection{Swimming velocity}
From the previous two  sections, the solute concentration distribution and the dual problem streamfunction are completely determined. The swimming velocity can now be computed using the reciprocal theorem, Eq.~\eqref{eq:reciprocal}. Because the phoretic slip $\ub_s$ is purely along $\eb_\mu$ and the normal unit vector to the sphere's surface is $\pm\eb_\tau$, we only need to compute $F^*$ and $\sigma_{\tau\mu}^* =\eb_\tau\cdot\grad\ub^*\cdot\eb_\mu+\eb_\mu\cdot\grad\ub^*\cdot\eb_\tau$.

Using Eq.~\eqref{eq:bcpsi}, on the boundaries of the spheres, we obtain 
\begin{equation}
\frac{a \sigma^*_{\tau\mu}(\tau_\pm,\mu)}{\sqrt{1-\mu^2}}
=\sum_{n=1}^\infty L_n'(\mu)S_n-\cosh\tau+\frac{\sinh^2\tau}{2(\cosh\tau-\mu)}\\
\end{equation}
with
\begin{align}
S_n &= -(\cosh\tau_\pm-\mu)^{3/2}U_n''(\tau_\pm)\nonumber\\
&+\frac{\sqrt{\cosh\tau_\pm-\mu}}{2}\left[\sinh\tau_\pm U_n'(\tau_\pm)+3\cosh\tau_\pm U_n(\tau_\pm)\right]
%-\frac{1}{a\sqrt{1-\mu^2}}\left[\pard{}{\tau}\left((\cosh\tau-\mu)^3\pard{\psi^*}{\tau}\right)-(1-\mu^2)\pard{}{\mu}\left((\cosh\tau-\mu)^3\pard{\psi^*}{\mu}\right)\right],\\
%-\frac{1}{a\sqrt{1-\mu^2}}\left[\sqrt{\cosh\tau-\mu}\pard{}{\tau}\left((\cosh\tau-\mu)^{5/2}\pard{\psi}{\tau}\right)+(1-\mu^2)\left(\cosh\tau-\frac{\sinh^2\tau}{2(\cosh\tau-\mu)}\right)\right],\nonumber\\
\end{align}

The total hydrodynamic force on the two-sphere system in the dual problem was computed in Ref.~\cite{stimson1926} as
\begin{align}
F^*=-\frac{4\pi\sqrt{2}}{a}\sum_{n=1}^\infty n(n+1)(\alpha_n+\gamma_n),
\end{align}
so that  the swimming velocity of the two-sphere system is obtained as
\begin{align}
U=\frac{2\pi a^2}{F^*}\bigg( &\int_{-1}^1\frac{u_\mu^s(\tau_+,\mu)\sigma^*_{\tau\mu}(\tau_+,\mu)\dd\mu}{(\cosh\tau_+-\mu)^2}\\
& \left. -\int_{-1}^1\frac{u_\mu^s(\tau_-,\mu)\sigma^*_{\tau\mu}(\tau_-,\mu)\dd\mu}{(\cosh\tau_--\mu)^2}\right).\label{eq:velocity}
\end{align}

In the previous equation, the integrals in $\mu$ are performed numerically (knowledge of the values of $a_n$, $b_n$, $\alpha_n$, $\beta_n$, $\gamma_n$ and $\delta_n$ completely determines $u_\mu^s$ and $\sigma^*_{\tau\mu}$ on the  boundary of the spheres).

\subsection{Results: self-propulsion of a two-sphere system}
The radius of the largest sphere is chosen as reference length scale, so that with no loss of generality we  set $R_1=1$ and $R_2<1$. In the following, the self-propulsion properties and their dependence on  $d$ and $R_2$ are examined. When $R_2=1$, the system is up/down-symmetric and there is no net motion. Similarly, when $R_2\ll 1$, the concentration distribution is only marginally impacted by the presence of the second sphere, and any net propulsion velocity is infinitesimal. We thus expect the presence of an optimal ratio of sphere sizes. 

\subsubsection{Far-field}
This is the result from the introduction, which we quote in dimensionless terms
\begin{equation}\label{eq:farfieldresult2}
\Ub=\frac{AM R_2(R_2-1)}{ d^2(R_2+1)}\eb_z.
\end{equation}

When $AM>0$ (particle with positive mobility releasing solute), in the far-field limit, the self-propulsion velocity is always oriented toward the smaller sphere. In this limit a maximum amplitude of the velocity is obtained for the optimal radius $R_2=\sqrt{2}-1\approx 0.41$ and is equal to $U_\textrm{max}=(\sqrt{2}-1)^2AM/d^2\approx 0.17 AM/d^2$.

\subsubsection{Velocity for arbitrary distances}
When the size of the spheres is no longer small compared to $d$, higher order corrections in both the distribution of solute and the hydrodynamic field can become significant, and even dominant in the limit where the contact distance, $d_c=d-(R_1+R_2)$ becomes small. 

\begin{figure}
\begin{center}
\begin{tabular}{cc}
\includegraphics[width=.45\textwidth]{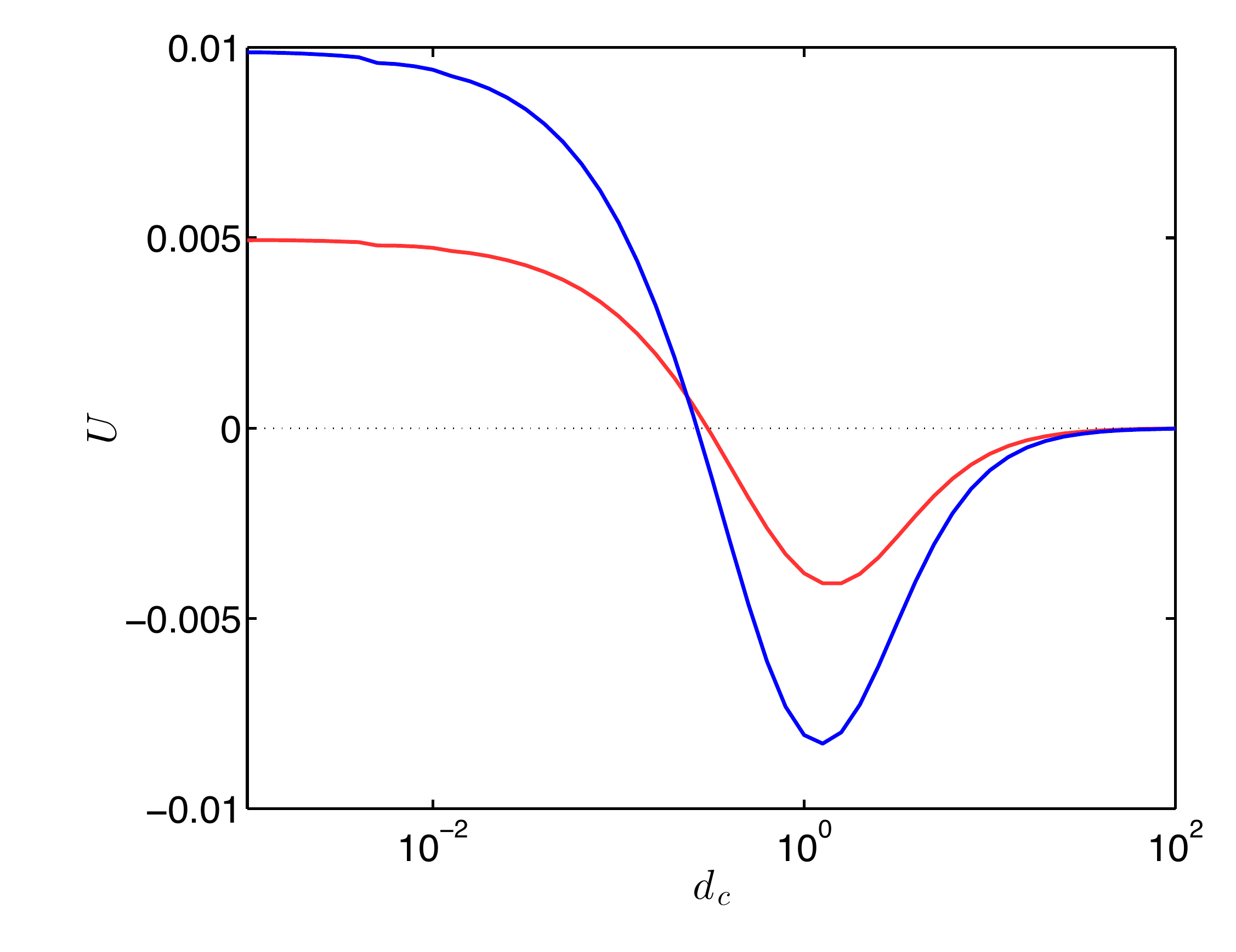}\\
\includegraphics[width=.45\textwidth]{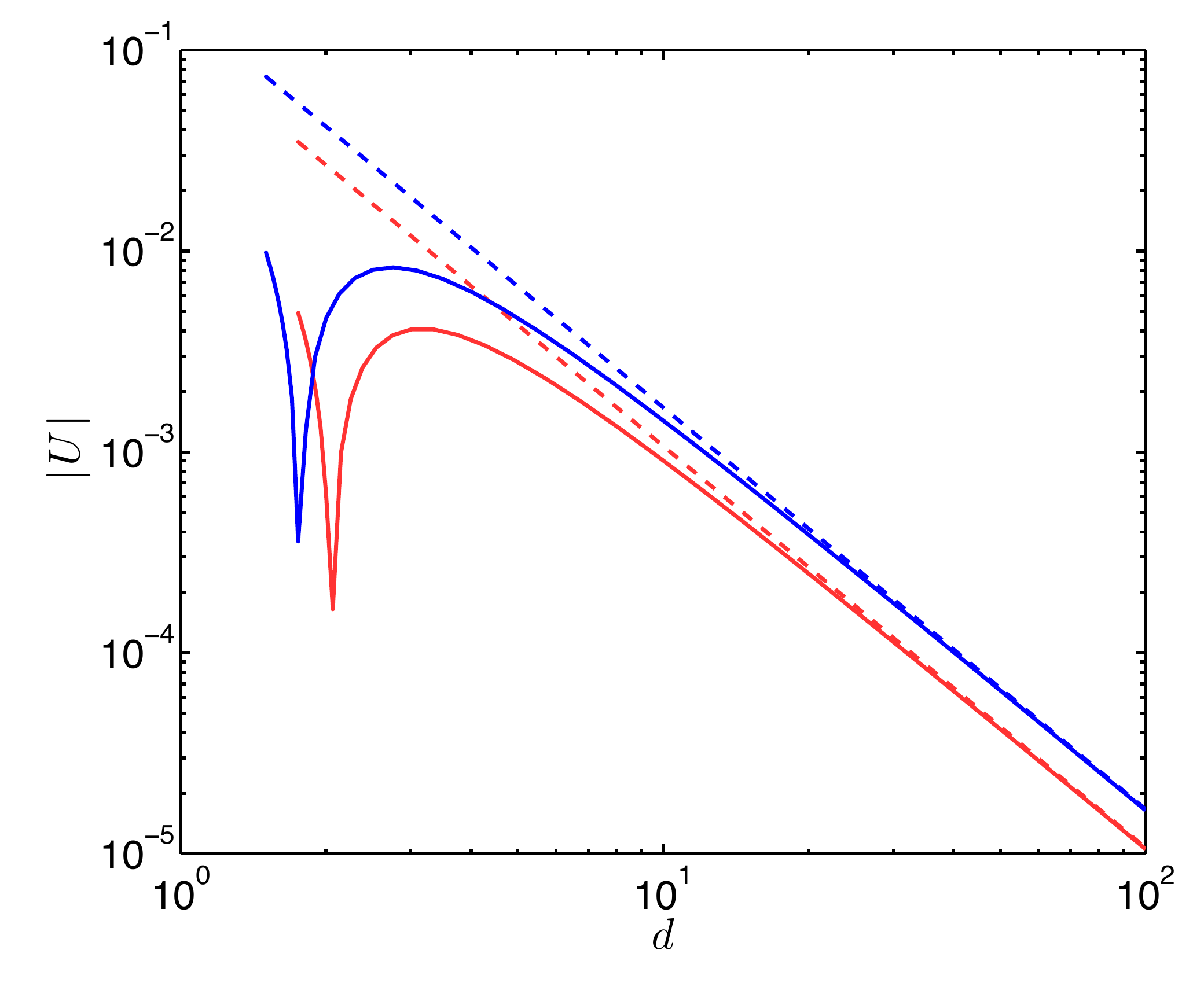} 
\end{tabular}
\caption{Top: Dependence  of the self-propulsion velocity with the contact distance between the two spheres, $d_c$, for $R_2/R_1=0.75$ (light red) and $R_2/R_1=0.5$ (dark blue). Bottom: Same quantities plotted in log-log scale to show the scaling and compare with the far-field predictions (dashed lines). Here, both activity $A=M=+1$.}\label{fig:evold}
\end{center}
\end{figure}

In Fig.~\ref{fig:evold} we plot   the dependence  of the swimming speed,  $U$, with $d_c$ when $R_2/R_1=0.5$ and $R_2/R_1=0.75$.  Strikingly,  the variation is non-monotonous  and the velocity even changes sign. Specifically, at a  small contact distance, the self-propulsion velocity is positive (larger sphere in front) while at larger distances (and in the far-field) the self-propulsion velocity is negative (smaller sphere in front). As a consequence there is a finite contact distance $d_c$ for which $U=0$ despite the asymmetry in the geometry of the system.

At large distance, the self-propulsion decreases as $d^{-2}$, a direct result of the decay of the concentration gradient created by each sphere on the other. This scaling is consistent with the far-field analysis, and the numerical result obtained from the complete calculation shows a good quantitative agreement with the far-field asymptotic solution obtained in the previous section, Eq.~\eqref{eq:farfieldresult2} (Fig~\ref{fig:evold}, bottom).

When the spheres are nearly in contact, $d_c\rightarrow 0$, the self-propulsion velocity converges to a finite and well-defined value, emphasizing that the concentration distribution and flow field within the narrowing gap only has marginal influence on the dynamics of the entire system. In that limit, the two-sphere system behaves  like a single solid with asymmetric (snowman) shape. \change{We note that obviously our results are only applicable while the thin-layer approximation at the core of the classical continuum framework remains valid (i.e.~sufficiently  large values of  $d_c$). However, because of the limited extent of the contact region  and its orientation relative to the direction of motion, we expect  the results to be only marginally modified by a correction taking into account the finite size of the interaction layer.}

\begin{figure}
\begin{center}
\includegraphics[width=9cm]{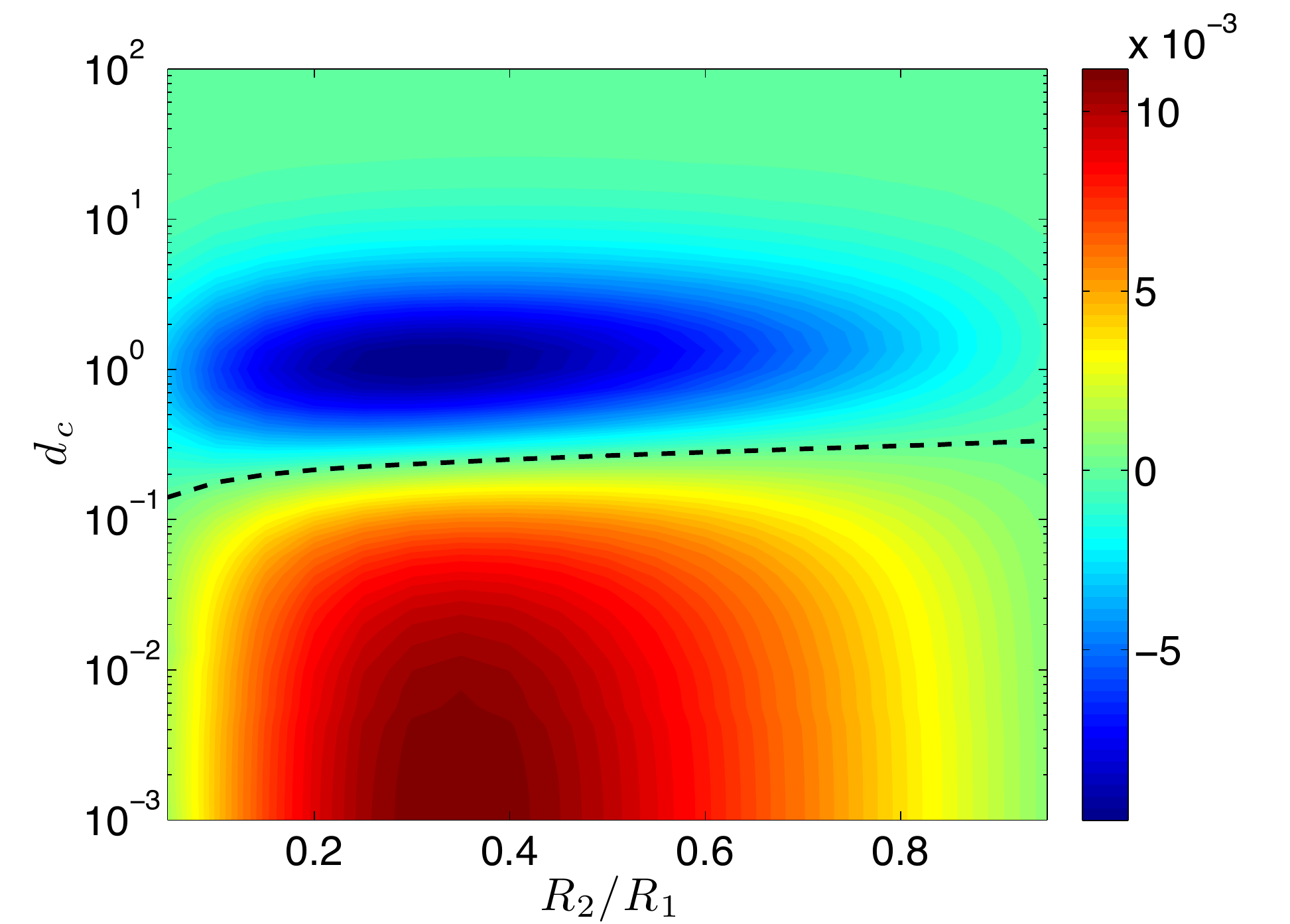}
\caption{Dependence  of the swimming velocity with the size ratio, $R_2/R_1$, and the closest distance, $d_c=d-(R_1+R_2)$, between the two spheres. The dashed line indicates the  configurations leading to no self-propulsion ($U=0$).}\label{fig:map}
\end{center}
\end{figure}

These results   can easily be generalized to arbitrary size ratio, $R_2/R_1$, as plotted in Fig.~\ref{fig:map}. Regardless of the relative radius of the two spheres,  self-propulsion occurs in different directions in the limit of near-contact and larger distances, and thus there always exist a finite contact distance $d_c\approx 0.1$--$0.2$ for which no propulsion is observed. As expected, and optimal propulsion velocity is obtained for intermediate values of the  size ratio (typically $R_2/R_1\approx 0.3$--$0.4$), whose dependence with $d_c$ is  weak.

Within the $(d_c, R_2/R_1)$-plane, two optimal configurations can be identified. The first one, which is the global maximum, corresponds to two spheres in contact with $R_2/R_1=0.35$ resulting in a velocity of $U= 0.011$. The second corresponds to a finite distance $d_c=1.15$ and a size ratio $R_2/R_1=0.31$, resulting in $U=-0.0098$.

\subsubsection{Solute distribution}
\begin{figure*}
\begin{center}
\includegraphics[width=.95\textwidth]{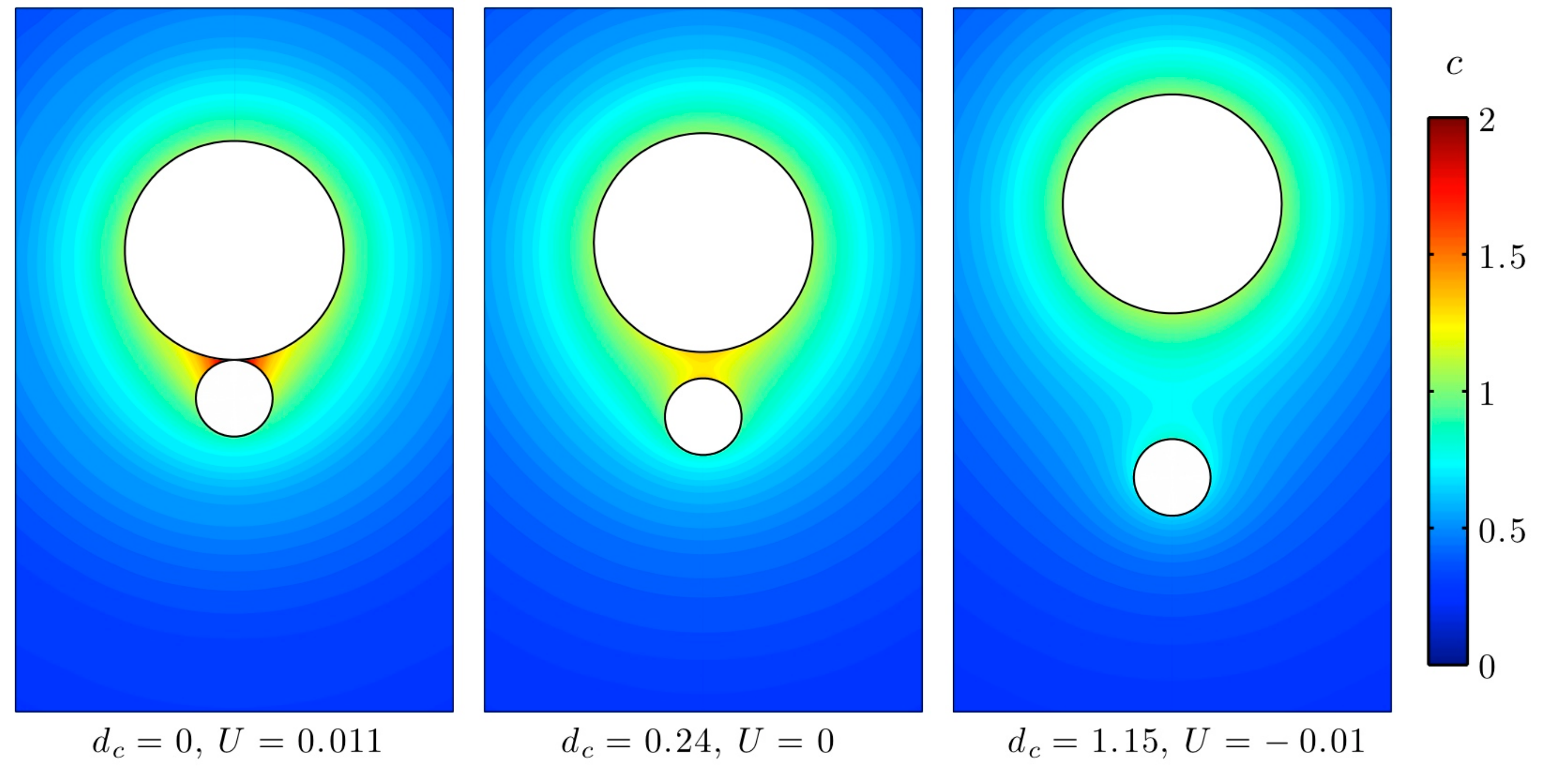} 
\caption{Solute concentration around a two-sphere system for a sized ratio $R_2/R_1=0.35$ and three contact distances: $d_c=0$  (left),  $d_c=0.24$ (center), and  $d_c=1.15$ (right). }\label{fig:concentration}
\end{center}
\end{figure*}

The  results above show that the distance between the two spheres critically impacts the propulsion velocity, particularly in determining its sign. To gain a better understanding of this effect, we show in Fig~\ref{fig:concentration} the solute concentration distribution for $R_2/R_1=0.35$ (maximum velocity) and increasing distance. When $A>0$ (solute release at the surface of the spheres), the solute concentration is always greater between the spheres than on the outside due to confinement: there the diffusive flux of solute can only take place on a reduced set of spatial directions, leading to an increase in the solute concentration and in the gradient. This effect is even more pronounced when the two spheres are in contact, leading to singular, but integrable, solute concentration gradients near the contact point.

Because of this solute distribution, the slip velocity is always oriented in opposite directions on the two spheres, regardless of their distance: when $AM>0$ (resp. $AM<0$), the slip velocity on each sphere is oriented toward (resp. away from) the other sphere. The contributions of each sphere to the propulsion are therefore always of opposite sign. In other words,  if the spheres were independent they  would move in opposite directions, or, perhaps more quantitatively, the contribution of both particles to Eq.~\eqref{eq:velocity} have opposite signs. 

\begin{figure}[t]
\begin{center}
\includegraphics[width=.45\textwidth]{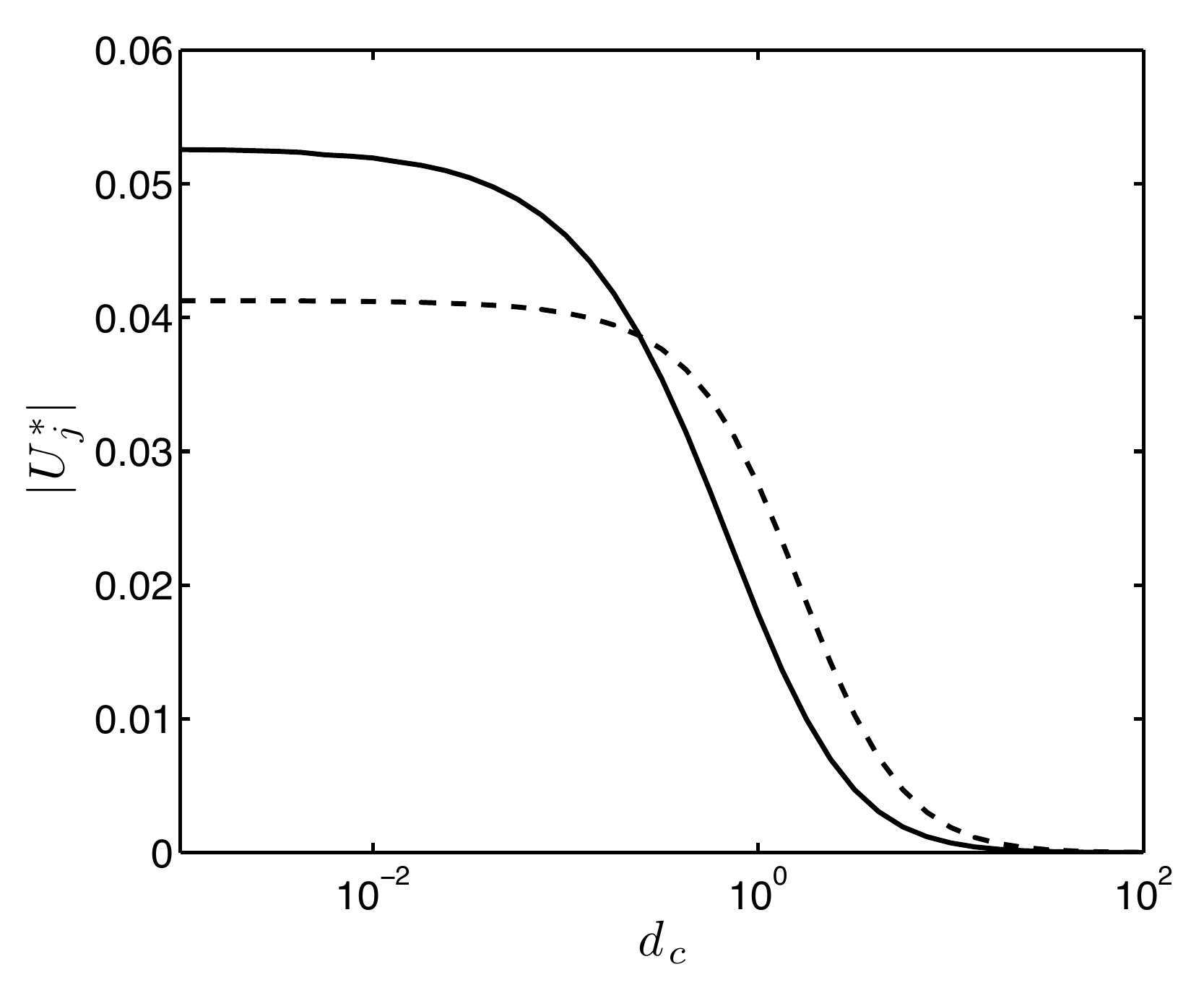}
\caption{Dependence  of the magnitude of the contribution of sphere 1 (solid) and 2 (dashed) to the self-propulsion velocity  (defined as the velocity of the two-sphere system if the mobility of the other sphere  is  zero) with  $d_c$. }\label{fig:contrib}
\end{center}
\end{figure}

In order to quantify this more precisely, we define $U_j^\dag$ with $j=1$ or $j=2$, the velocity of the two-sphere system obtained when only sphere $j$ has non zero mobility (the other sphere's mobility is set to zero). By linearity, the real self-propulsion velocity is obtained as $U=U_1^\dag+U_2^\dag$, and $U_j^\dag$ can be seen as a measure of the contribution to the total self propulsion. We plot in Fig.~\ref{fig:contrib}   the dependence of  the magnitude of $U_1^\dag$ and $U_2^\dag$ with $d_c$ (note that we always have  $U_2^\dag<0<U_1^\dag$). It shows, indeed, that rather than a fundamental change in the contribution of each sphere to the propulsion, the relative variations (and the slower decay of the contribution of the smaller sphere) is responsible for the change of sign in propulsion velocity. 

More specifically, for small $d_c$, the contribution of the largest sphere dominates, a consequence from the large concentration gradients generated on that sphere near the contact point by the presence of the smaller one (see Fig.~\ref{fig:concentration}, left). This effect, mainly due to confinement, reduces rapidly as $d_c$ increases, and the surface concentration distribution on the larger sphere loses its strong asymmetry. In contrast, the asymmetry of the concentration distribution on the smaller sphere is maintained at larger distance (Fig.~\ref{fig:concentration}, right).

\subsubsection{Reactive vs. diffusive effects}

The chemical reaction at the surface of the spheres addressed so far is a simple fixed-flux release/absorption of solute. A more general one-step reaction  can be considered by assuming that the solute is consumed with a uniform reaction rate, so that the dimensionless boundary condition on each sphere is replaced by \citep{michelin2014}
\begin{equation}
\nb\cdot\grad c=1+\Da c.
\end{equation}
Here the Damk\"ohler number $\Da=\mathcal{K}a/\kappa$ is a measure of the relative magnitude of reaction and diffusion. Note that $\Da=0$ corresponds to the previous situation (with $A=-1$). \change{A generalization of this one-step fixed-rate approach to chemical release ($\Da<0$) would not be physically relevant as the release rate would be proportional to the local concentration, leading to a local exponential increase of the concentration and thus no steady solution to the diffusive problem. We exclusively focus on $\Da>0$ in the following.}

\begin{figure}
\begin{center}
\includegraphics[width=.45\textwidth]{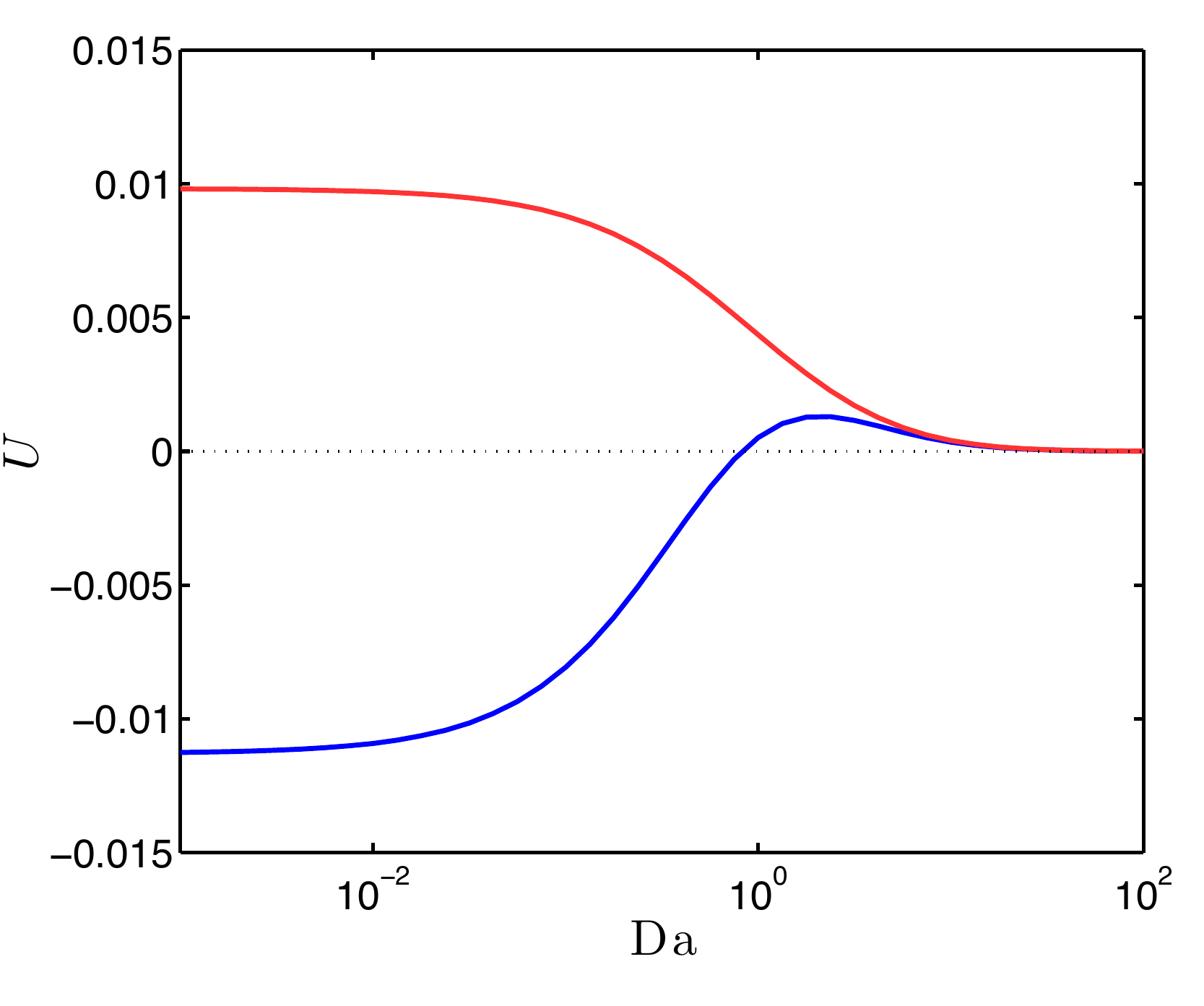}
\caption{Dependence of the self-propulsion velocity of the two-sphere system  with $\Da$ for the two optimal configurations for  $\Da=0$: $R_2/R_1=0.35$ and $d_c=0$ (dark blue), and $R_2/R_1=0.31$ and $d_c=1.15$ (light red). The results of the previous section are recovered when $\Da=0$ provided that the velocity sign is reversed ($A=-1$  here).}\label{fig:Daeffect}
\end{center}
\end{figure}

In Fig.~\ref{fig:Daeffect} we plot  the evolution with $\Da$ of the swimming velocity for the two optimal configurations identified previously. In both cases, reactive effects are observed to significantly reduce the magnitude of the propulsion velocity and $U\sim \Da^{-2}$ for $\Da\gg 1$. As  discussed in Ref.~\cite{michelin2014}, for finite $\Da$, the solute consumption is limited by the reduction of its local concentration, which tends to reduce concentration gradient and slip velocities. 

  For the optimal configuration where both spheres are in contact,  a surprising change in the sign of the velocity is observed at finite values of $\Da$. While the two-sphere system self-propels in the direction of the smaller sphere at small $\Da$ (fixed-flux absorption, remembering that this is equivalent to the results of Fig.~\ref{fig:map} when correcting for the change of sign of the activity), finite-$\Da$ effects lead to a propulsion velocity in the direction of the larger sphere. 

\begin{figure*}
\begin{center}
\begin{tabular}{cc} 
\includegraphics[height=6.5cm]{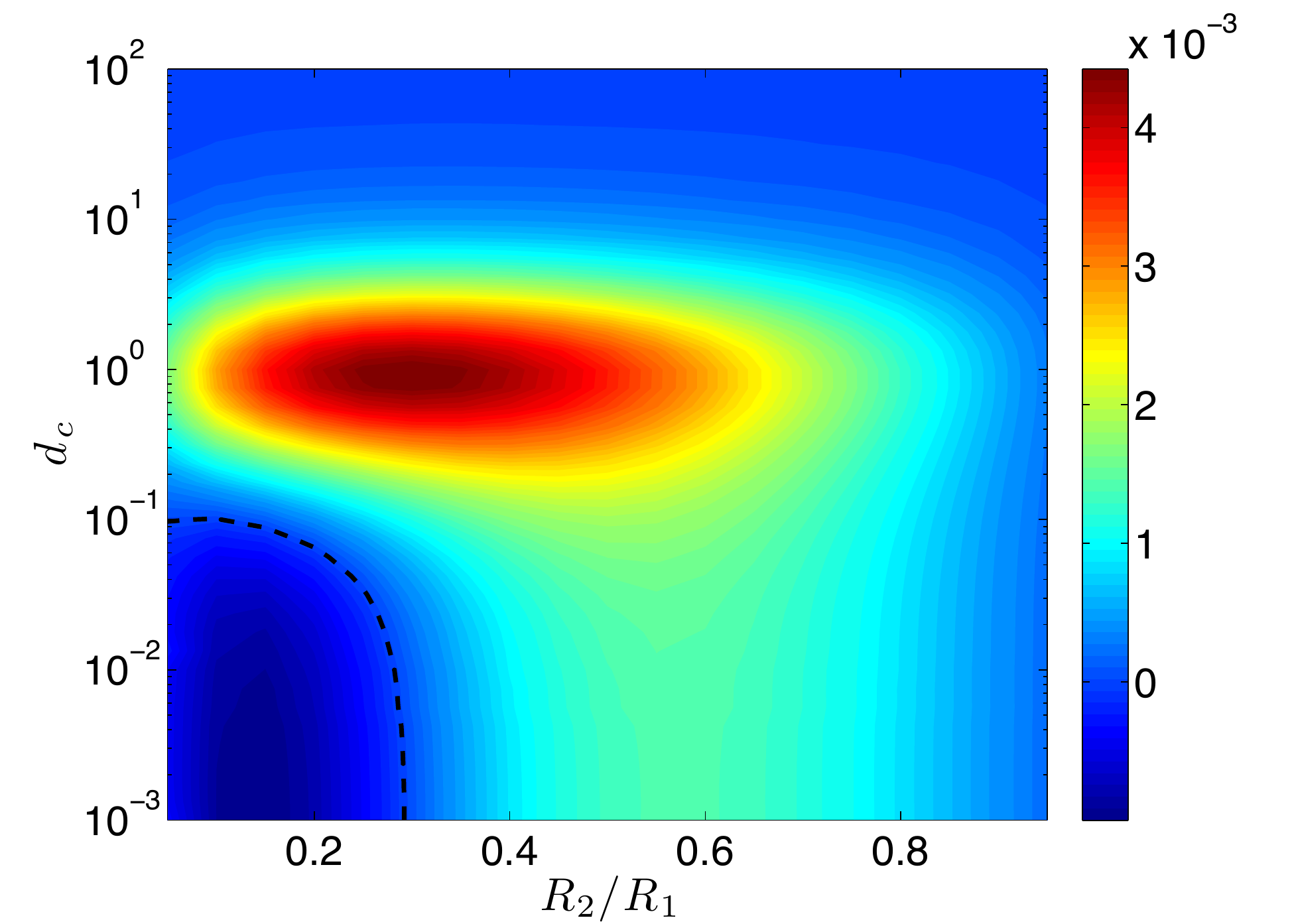}&
\includegraphics[height=6.5cm]{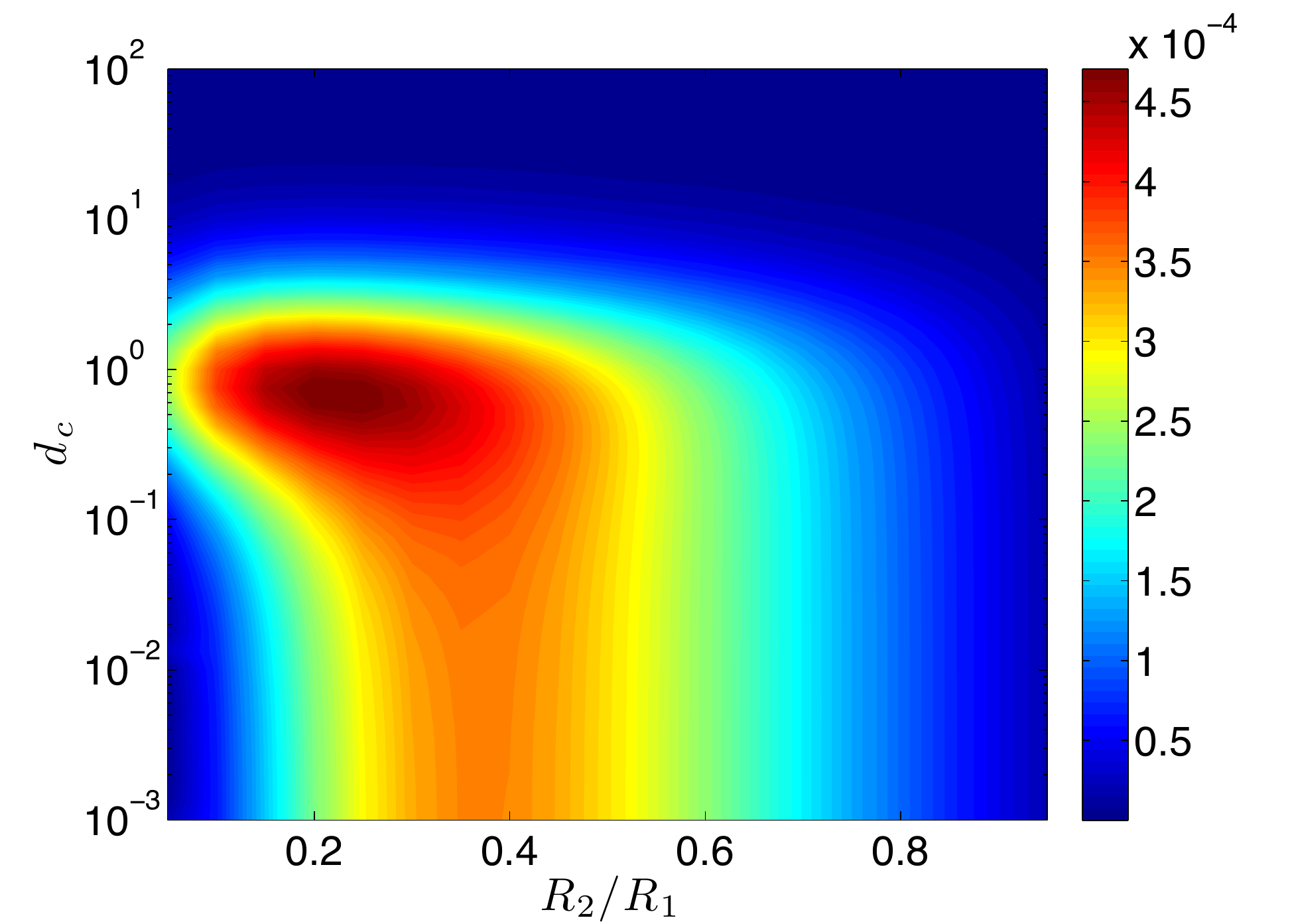}
\end{tabular}
\caption{Same as Figure~\ref{fig:map} for $\Da=1$ (left) and $\Da=10$ (right). Since $A=-1$, the swimming direction is reversed.}\label{fig:map_Da}
\end{center}
\end{figure*}

This modification of the propulsion properties of the system for near-contact configuration is confirmed on Fig.~\ref{fig:map_Da} where we plot similar results to those from Fig.~\ref{fig:map}  for $\Da=1$ (left) and $\Da=10$ (right). When $d_c=O(1)$, the effect of $\Da$ is a global reduction in the propulsion velocity (note the change of scales between Figs.~\ref{fig:map} and \ref{fig:map_Da}). However, when the spheres are close to each other, or even touching, the direction of propulsion is reversed in comparison to the case $\Da=0$. This effect occurs for smaller $\Da$ when the contrast between the sizes of the spheres is smaller. As a result, the existence of a non-propelling configuration for each size ratio is lost at finite and large $\Da$, and  most of the two-sphere swimmers self-propel in the same direction (namely with the larger sphere at the front), the direction being dictated by the slip velocity on the smaller sphere when $A<0$: when reactive effects are significant, i.e. for $\Da=O(1)$, they limit the role of confinement in setting the concentration level and concentration gradients between the two spheres, thereby significantly reducing the contribution of the slip velocity on the larger sphere to the global motion.

%%%%%%%%%%%%%%%%%%%%%%
\section{Autophoretic locomotion of a homogeneous near-sphere}
\label{sec:near_sphere}
The previous section focused on a specific geometry, a system made of two spheres, which could be solved exactly even for large geometric  asymmetries. We now turn to a different limit in which we address arbitrary perturbations in the shape of a spherical colloidal particle. In order to be able to compute the influence of each surface mode on the locomotion, we have to assume that the amplitudes of the shape perturbations are small compared to the typical length scale of the particle. This problem is similar to that presented by Ref.~\cite{shklyaev2014}, but using a different calculation framework.

\subsection{Geometry definition}
We consider here an axisymmetric particle of uniform surface properties, whose surface can be described in spherical polar coordinates as $R(\mu)=1+\e\xi(\mu)$, with $\mu=\cos\theta$ where $\theta$ is the polar angle. We investigate the near-sphere limit, namely $\e\ll 1$, and consider the  general case of a one-step kinetic reaction at the surface so that the activity and mobility of the particle are described by 
\begin{equation}
\nb\cdot\grad c=1+\Da\, c,\quad \ub=(1-\nb\nb)\cdot\grad c,\label{eq:bc_adim}
\end{equation}
at $r=1+\e\xi$, and $\nb$ is the normal unit vector pointing into the fluid domain
\begin{equation}
\nb=\frac{\eb_r+\e\xi'\sqrt{1-\mu^2}\eb_\theta}{\sqrt{1+\e^2(1-\mu^2)\xi^{'2}}}.\label{eq:normal}
\end{equation}

We seek a solution of the autophoretic propulsion problem as a regular series expansions in $\e\ll 1$ of the solute concentration and velocity fields, and aim to compute the leading order contribution to the propulsion velocity
\begin{subeqnarray}\label{eq:expansion}
c&=&c^0+\e c^1+\e^2c^2+...\\
\ub&=&\ub^0+\e \ub^1+\e^2\ub^2+...\\
\Ub&=&\Ub^0+\e \Ub^1+\e^2\Ub^2+...
\end{subeqnarray}
Using Eq.~\eqref{eq:normal}, the expansion can also be applied to the  unit vector normal  to the surface of the particle
\begin{equation}
\nb=\nb_0+\e\nb_1+\e^2\nb_2+...,
\end{equation}
with 
\begin{subeqnarray}
\nb_0&=&\eb_r,\\
\nb_1&=&\xi'(\mu)\sqrt{1-\mu^2}\eb_\theta,\\
\nb_2&=&-\frac{\xi'(\mu)^2(1-\mu^2)}{2}\eb_r-\xi(\mu)\xi'(\mu)\sqrt{1-\mu^2}\eb_\theta.
\quad\end{subeqnarray}

The deformation of the particle radius, $\xi(\mu)$, is projected onto orthogonal Legendre polynomials, so we write $\xi(\mu)=\sum\xi_pL_p(\mu)$. Note that by definition of the mean particle's radius, $\xi_0$ must be zero. Also, at leading order $\xi_1$ only corresponds to a translation of the particle, its shape remaining symmetric. Hence, we do not expect any contribution from that mode to  self-propulsion, at least at leading order.

\subsection{Solute concentration problem}\label{sec:3.2}
At all orders in $\e$, the solute concentration $c^j(r,\mu)$ satisfies Laplace's equation so that, enforcing the far-field decay of $c$ we obtain
\begin{equation}
c^j(r,\mu)=\sum_{p=0}^\infty \frac{c_p^j L_p(\mu)}{r^{p+1}},
\end{equation}
where $c_p^j$ are constants obtained from the  boundary condition at the surface of the  particle, $\nb\cdot\grad c=1+\Da c$, which can be rewritten at $r=1$ using domain perturbations as
\begin{subeqnarray}
\pard{c^0}{r}-\Da c^0&=&1,\label{eq:bc1_0}\\
\pard{c^1}{r}-\Da c^1&=&-\xi\pard{^2c^0}{r^2}+\xi'(1-\mu^2)\pard{c^0}{\mu}+\Da \xi\pard{c^0}{r},\label{eq:bc1_1}\\
\pard{c^2}{r}-\Da c^2&=&\xi'(1-\mu^2)\pard{c^1}{\mu}+\frac{\xi^{'2}(1-\mu^2)}{2}\pard{c^0}{r}-\xi\pard{^2c^1}{r^2}\nonumber \\
&&-\xi\xi'(1-\mu^2)\pard{c^0}{\mu}+\xi\xi'(1-\mu^2)\pard{^2c^0}{r\partial\mu}\nonumber \\
&&-\frac{\xi^2}{2}\pard{^3c^0}{r^3}+\Da \left(\xi\pard{c^1}{r}+\frac{\xi^2}{2}\pard{^2c^0}{r^2}\right)\label{eq:bc1_2},
\end{subeqnarray}
where all quantities  are evaluated at $r=1$.

At leading order, the solution to the spherical problem  is trivially obtained for $c^0$ as the isotropic solution
\begin{equation}
c^0=-\frac{1}{(1+\Da)r}\cdot\label{eq:c0}
\end{equation}
After substitution of this result into Eq.~\eqref{eq:bc1_1}, the first-order correction to the isotropic solution concentration is obtained as $c^1_p=2A\xi_p/(p+1)$ or equivalently
\begin{equation}
c^1(r,\mu)=-\frac{2+\Da}{1+\Da}\sum_{p=0}^\infty\frac{\xi_pL_p(\mu)}{(p+1+\Da)r^{p+1}}\cdot\label{eq:c1}
\end{equation}

Finally, Eq.~\eqref{eq:bc1_2} leads to the following solution for the second-order correction $c^2$ 
\begin{equation}
c^2(r,\mu)=-\frac{1}{1+\Da}\sum_{m,n,p=0}^\infty\frac{E_{mnp}\xi_m\xi_nL_p(\mu)}{(p+1+\Da)r^{p+1}},\label{eq:c2}
\end{equation}
where the third-order tensor $E_{mnp}$ is defined as 
\begin{eqnarray}
%E_{mnp}=\left(A_{mnp}+\frac{n(n-3)}{2(2n+1)}B_{mnp}\right),\\
E_{mnp}&=&\frac{2p+1}{2}\int_{-1}^1L_pQ_{mn}\dd \mu\label{eq:emnp}\\
Q_{mn} &=& \left[\frac{(2n+1)(n+1)+n(n+4)\Da+n\Da^2}{n+1+\Da}\right]L_mL_n\nonumber\\
&&+\frac{(n-3-\Da)}{2(n+1+\Da)}(1-\mu^2)L_m'L_n',
\end{eqnarray}
 where we note 
  $L_p\equiv L_p(\mu)$.

\subsection{Stokes flow and swimming problems}
The mobility  of the particle, Eq.~\eqref{eq:bc_adim}, imposes a tangential  forcing on the fluid outside the particle resulting in a global flow field in the Stokes regime. Using Eqs.~\eqref{eq:expansion}, this boundary condition expressed at $r=1+\e\xi$ can be converted, at each order, into a condition on the flow velocity on the spherical boundary at $r=1$
\begin{subeqnarray}
\ub^0&=&-M\sqrt{1-\mu^2}\pard{c^0}{\mu}\eb_\theta,\label{eq:bc2_0}\\
\ub^1&=&-\xi\pard{\ub^0}{r}-M\sqrt{1-\mu^2}\left(\pard{c^1}{\mu}+\xi\pard{^2c^0}{r\partial\mu}\right)\eb_\theta\nonumber\\
&&
+M(1-\mu^2)\xi'\pard{c^0}{\mu}\eb_r-M\xi'\sqrt{1-\mu^2}\pard{c^0}{r}\eb_\theta,\label{eq:bc2_1}\\
\ub^2&=&-\xi\pard{\ub^1}{r}-\frac{\xi^2}{2}\pard{^2\ub^0}{r^2}\nonumber\\* 
&&-M\sqrt{1-\mu^2}\left(\pard{c^2}{\mu}+\xi\pard{^2c^1}{r\partial\mu}+\frac{\xi^2}{2}\pard{^3c^0}{r^2\partial\mu}\right)\eb_\theta\nonumber\\* 
&&+M(1-\mu^2)\xi'\left(\pard{c^1}{\mu}+\xi\pard{^2c^0}{r\partial\mu}\right)\eb_r\nonumber\\*
&&
-M\xi'\sqrt{1-\mu^2}\left(\pard{c^1}{r}+\xi\partial{^2c^0}{r^2}\right)\eb_\theta\nonumber\\
&& +M\xi^{'2}(1-\mu^2)\left(\pard{c^0}{r}\eb_r+\sqrt{1-\mu^2}\pard{c^0}{\mu}\eb_\theta\right)\nonumber\\
&&+M\xi\xi'\sqrt{1-\mu^2}\left(\pard{c^0}{r}\eb_\theta-\sqrt{1-\mu^2}\pard{c^0}{\mu}\eb_r\right).\quad\quad\label{eq:bc2_2}
\end{subeqnarray}
where all  quantities  are again to be computed at $r=1$. 

We now have to  solve for the flow outside the unit sphere subject, at each order, to the boundary conditions at $r=1$ and  to the constraint that the total force on the unit sphere is exactly zero. This is strictly equivalent to solving for the flow outside the force-free non-spherical particle since the flow is also force-free ($\nabla\cdot\boldsymbol\sigma=0$) and 
\begin{equation}
\int_{r=1}\!\!\!\!\!\!\boldsymbol\sigma\cdot\nb\dd S-\int_{r=1+\xi}\!\!\!\!\!\!\!\!\!\!\!\!\!\boldsymbol\sigma\cdot\nb\dd S=\int_{V_1}\!\!\!\nabla\cdot\boldsymbol\sigma\dd V-\int_{V_2}\!\!\!\nabla\cdot\boldsymbol\sigma\dd V=0,
\end{equation}
where $V_1$ (resp. $V_2$) is the domain located inside (resp. outside) the unit sphere and outside (resp. inside) the actual particle.

Replacing the flow problem on the unit sphere is particularly convenient as it gives access to two major analytical tools of low-$\Rey$ swimming problems: (i) the reciprocal theorem to determine the swimming velocity $\Ub$ from the flow velocity on the unit sphere $\ub_s$, Eqs.~\eqref{eq:bc2_0}, and (ii) the squirmer framework that provides an analytic solution for the flow velocity everywhere by projecting boundary conditions  onto orthogonal squirming modes \citep{blake1971}. In particular, the reciprocal theorem applied to a unit sphere provides the swimming velocity at each order from the surface velocity distribution as \citep{stone1996}
\begin{equation}
\Ub^j=-\mean{\ub^j_s}=\frac{\eb_z}{2}\int_{-1}^1\left(u_\theta^j\sqrt{1-\mu^2}-u^j_r\mu\right)\dd\mu.\label{eq:recthm}
\end{equation}

In \S \ref{sec:3.2}, we obtained that $c^0$ was isotropic. Consequently, it does not create any flow at leading order, as expected, so that $\ub^0=0$. At $O(\e)$, the boundary condition on $\ub^1$ simplifies into
\begin{eqnarray}
\label{eq:u1} \ub^1&=&-M\sqrt{1-\mu^2}\pard{c^1}{\mu}\eb_\theta-M\xi'\sqrt{1-\mu^2}\pard{c^0}{r}\eb_\theta\\
&=&\frac{M}{1+\Da}\sqrt{1-\mu^2}\left[\sum_{p=0}^\infty\left(\frac{1-p}{p+1+\Da}\right)\xi_pL_p'(\mu)\right]\eb_\theta\nonumber.
\end{eqnarray}

From Eq.~\eqref{eq:recthm} and using  Eqs.~\eqref{eq:c0}, \eqref{eq:c1} and \eqref{eq:bc2_1}, we obtain $\Ub^1=0$; hence, there is no net motion of the particle at $O(\e)$. However, a non-zero $O(\e)$ flow field is created through phoretic effects, and $\ub^1$ should be computed everywhere in order to obtain $\ub^2$ from Eq.~\eqref{eq:bc2_2}. At order $O(\e)$ the surface velocity is purely tangential. Using the squirmer model framework \citep{blake1971,michelin2011}, the flow field $\ub^1$ satisfying Stokes' equations and the force-free condition can be written as
\begin{align}
\nonumber \ub^1(r,\mu)=\sum_{n=1}^\infty& (2n+1)\alpha_n^1\left[\frac{\psi_n(r)}{r^2}L_n(\mu)\eb_r\right.\\
&\left.-\frac{1}{n(n+1)}\frac{\psi'_n(r)}{r}\sqrt{1-\mu^2}L_n'(\mu)\eb_\theta\right],\label{eq:u1_full}
\end{align}
with $\psi_n(r)$ the radial streamfunction for the $n$th squirming mode
\begin{equation}
\psi_1(r)=\frac{1-r^3}{3r},\,\,\, \psi_n(r)=\frac{1}{2}\left(\frac{1}{r^n}-\frac{1}{r^{n-2}}\right)\,n\geq 2,
\end{equation}
and $\alpha_n^1$, the amplitude of that mode, obtained from the surface velocity as
\begin{eqnarray}
\nonumber\alpha_n^1&=&\frac{1}{2}\int_{-1}^1\sqrt{1-\mu^2}L_n'(\mu)(\ub^1\cdot\eb_\theta)\dd\mu\\
&=&-\frac{M}{1+\Da}\frac{\xi_n n(n-1)(n+1)}{(2n+1)(n+1+\Da)}\cdot\label{eq:alphan}
\end{eqnarray}
The squirming modes intensities characterize, among other things, the swimming properties and the far-field velocity created by the particle; $\alpha_1=U$ is the particle's swimming velocity, and the first mode also include a potential source dipole contribution. The second mode is related to the slowest-decaying singularity in the far-field of a force-free and torque-free particle, namely a  stresslet \citep{batchelor1970}. The result in Eq.~\eqref{eq:alphan} confirms that there is no net swimming motion at first order ($\alpha_1^1=0$), but it however shows that the dominant stresslet is $O(\e)$ and dictated by the second Legendre mode in $\xi(\mu)$. A pure mode-2 shape change would then lead to no net propulsion, by symmetry, but to a non-zero force dipole which will impact the bulk stress  \citep{batchelor1970} and  hydrodynamic interactions in a suspension.

 The sign of $\alpha_2^1$ dictates the type of stresslet. When $\alpha_2^1>0$ this corresponds to puller systems where the thrust center is located in front of the drag center, similarly to the  flagellated alga  \emph{Chlamydomonas}; in contrast $\alpha_2^1<0$ correspond to pushers where the position of drag and force centers are reversed, as is the case for most flagellated bacteria.  For $M>0$ (resp. $M<0$), prolate particles ($\xi_2>0$) will act as pushers (resp. pullers) while oblate particles ($\xi_2<0$) will act as pullers (resp. pushers). As the solute consumption is enhanced near the poles of the prolate spheroid, when $M>0$ the slip velocity is oriented away from the pole and toward the equator (puller), while for $M<0$, the slip velocity is oriented from the equator to the poles (pusher). Interestingly,  this dominant flow field decays rapidly with $\Da$, a direct consequence of the slower diffusion: the reaction kinetics are slowed near the surface due to the depleted solute concentration resulting in smaller concentration gradients and slip velocities.

From Eqs.~\eqref{eq:u1_full} and \eqref{eq:alphan}, the velocity gradient $\partial \ub^1/\partial r$ at the surface of the sphere is now obtained as
\begin{align}
\pard{\ub^1}{r}(r=1)=&\frac{M}{1+\Da}\sum_{n=0}^\infty\xi_n n(n-1)\times\nonumber\\
&\hspace{.8cm}\left[L_n(\mu)\eb_r+\frac{2\sqrt{1-\mu^2}}{n+1}L_n'(\mu)\right].
\end{align}

Substitution into Eq.~\eqref{eq:bc2_2} leads, for the velocity on the boundary, to
\begin{equation}
\ub^2=\frac{M}{1+\Da}\sum_{m,n=0}^\infty\xi_m\xi_n \left(Y_{mn} \eb_r + Z_{mn} \sqrt{1-\mu^2}\eb_\theta\right),
\end{equation}
with
\begin{subeqnarray}
Y_{mn}&=&-\frac{n(n-1)(n+1)}{n+1+\Da}L_m(\mu)L_n(\mu)\nonumber\\
&& +\frac{n-1}{n+1+\Da}(1-\mu^2)L_m'(\mu)L_n'(\mu),\\
Z_{mn}&=&-\frac{(n-1)(2n-1+\Da)}{n+1+\Da}L_m(\mu) L'_n(\mu)\nonumber\\
&& -\frac{(n+1)(2+\Da)}{n+1+\Da}L_m'(\mu)L_n(\mu)\nonumber\\
&&+\sum_{p=0}^\infty E_{mnp}\frac{L'_p(\mu)}{p+1+\Da} \label{eq:u2bc}\cdot
\end{subeqnarray}
From the previous equation, the reciprocal theorem, Eq.~\eqref{eq:recthm}, can be used to compute $\Ub^2$ (see Appendix \ref{secB}) and at leading order we finally obtain $O(\varepsilon^2)$ locomotion  as
\begin{align}
\Ub&=\e^2\eb_z\sum_{n=0}^\infty a_n(\Da)\xi_n\xi_{n+1}+O(\e^3),\label{eq:U2}
\end{align}
with
\begin{align}
a_n(\Da)&=\frac{[(n-2)(n+2)-(n+7)\Da-3\Da^2]\Gamma_n}{(1+\Da)(2+\Da)(n+1+\Da)(n+2+\Da)},\\
\Gamma_n&=\frac{2Mn(n+1)(n-1)}{(2n+1)(2n+3)}.
\end{align}
After having accounted for the difference in dimensional reference  velocity, Eq.~\eqref{eq:U2} is exactly equivalent  to the result recently derived in Ref.~\citep{shklyaev2014} using a different, colloidal calculation framework.  

This final  result shows that it is indeed possible to create self-propulsion through shape asymmetries.  In the case of infinitesimal perturbations of a spherically homogenous autophoretic particle, the resulting swimming velocity is quadratic in the perturbation amplitude. The form of the result, Eq.~\eqref{eq:U2}, is consistent with the fact that $U^2$ is zero for front-back symmetric particles (i.e. those with only even Legendre modes in their shape function $\xi$). One may also notice that as expected $\xi_0$ and $\xi_1$ do not contribute at this order as they merely change the radius or the center position of the sphere, not modifying its isotropy. More surprising is the role played by the second mode of deformation $\xi_2$ that dictates the dominant stresslet at $O(\e)$: when $\Da=0$ (fixed-flux emission/absorption), such a deformation does not contribute to the leading order propulsion, while it does as soon as $\Da\neq 0$. 

%\change{We show here that for purely diffusive solute, self-propulsion of a near-sphere only occur at $O(\e^2)$. When solute advection by phoretic flows is taken into account, this result may however be modified (see Ref.~\cite{khair2013} for the case of a near-sphere in an externally-imposed chemical field).}
% Seb - du coup j'ai vire

When $n\gg 1$, we notice that $a_n=O(n)$, and therefore, the infinite sum in Eq.~\eqref{eq:U2} only converges if $\xi_n=o(1/n)$ when $n\gg 1$, a condition which is  satisfied for regular shapes. If $\gamma$ denotes the angle between $\nb$ and $\eb_r$, then by definition of $\nb$ we have
\begin{equation}\label{eq:gammadef}
\tan\gamma=R'(\mu)\sqrt{1-\mu^2}.
\end{equation}
It can be shown that a condition for divergence of the sum in Eq.~\eqref{eq:U2} is  the existence of a non-integrable singularity in $\tan\gamma$. This, however, only includes marginal cases where the perturbative framework fails in the development of $\nb$ as a regular perturbation series.

A final interesting observation arises regarding the sign of the propulsion speed at different values of $\Da$. From Eq.~\eqref{eq:U2},
we observe that for all $n$, the product $a_n(\Da\ll 1)\cdot a_n(\Da\gg 1)$ is always negative. In other words,  for {any} shape (i.e. any coefficients $\xi_n$), the particle will swim in opposite direction in the diffusion-dominated regime ($\Da\gg1$) and reaction-dominated regime ($\Da\ll 1$). %The origin of this direction-reversal is a complex combination of the modification of concentration field near the perturbed surface and of the slip velocity distribution at second order.

\subsection{Optimal autophoretic near-sphere}
We now use our asymptotic result, Eq.~\eqref{eq:U2}, in order to compute the optimal way to distribute the surface perturbation modes  maximizing the magnitude of the swimming velocity. 

\subsubsection{Optimization framework}
Formally, the leading order swimming velocity, Eq.~\eqref{eq:U2}, can be expressed in terms of the vector $\xib=\e(\xi_n)_n$ as a bilinear form
\begin{equation}
U=\xib\cdot\mathbf{K}(\Da)\cdot\xib,
\end{equation}
with the upper-diagonal linear operator $\mathbf{K}$ defined as
\begin{equation}
K_{mn}(\Da)=a_n(\Da)\delta_{m,n+1}.
\end{equation}
Since the swimming velocity is quadratic in the amplitude of the perturbation, optimization requires to impose some kind of fixed norm to guarantee that we remain within the perturbative framework in which this result was obtained. Hence, we define an objective function $J$
\begin{equation}
J=\xib\cdot\mathbf{K}\cdot\xib-\lambda\left(\frac{1}{2}\xib\cdot \mathbf{H}\cdot \xib-1\right),\label{eq:evp}
\end{equation}
with $\lambda$ a Lagrangian multiplier and $\mathbf{H}$ a bilinear, symmetric positive definite operator corresponding to the particular norm chosen for $\xi$. Maximizing or minimizing $U$ is therefore equivalent to seeking solutions of
\begin{equation}
\left(\mathbf{K}+\mathbf{K}^T\right)\cdot\xib=\mathbf{H}\cdot\xib,\qquad \textrm{with\,\,\,\,   }\frac{1}{2}\xib\cdot \mathbf{H}\cdot \xib=1,
\end{equation}
The previous equation is effectively an eigenvalue problem for $(\xib,\lambda)$ with a constraint on the norm of the eigenvectors $\xib$. In practice, for a given norm operator $\mathbf{H}$, we seek the solutions of Eq.~\eqref{eq:evp} for a finite number $N$ of modes in $\xi_n$. This leads to a set of $N$ numerical solutions, from which the shape with maximum velocity is extracted.

The choice of the norm used to define $\mathbf{H}$  significantly impacts, as expected, the convergence of the optimal solution when $N\rightarrow\infty$. For example, a simple $L_2$-norm of the perturbation amplitude $\xi$ (denoted $\mathbf{H}^{(1)}$) or of local slope perturbation (i.e. angle between $\nb$ and $\eb_r$, denoted $\mathbf{H}^{(2)}$), which  led, respectively,  to the following tensors
\begin{eqnarray}
H_{mn}^{(1)}&=&\int_{-1}L_m(\mu)L_n(\mu)\dd\mu=\frac{2}{2n+1}\delta_{mn},\\
 H_{mn}^{(2)}&=&\int_{-1}^1(1-\mu^2)L_m'(\mu)L'_n(\mu)\dd\mu=\frac{2n(n+1)}{2n+1}\delta_{mn},\quad
\quad\end{eqnarray} 
do not guarantee the convergence of the result as $N\rightarrow \infty$. In fact, in both cases, for finite values of $N$, the optimal shape is obtained as a sharp cusp near the pole leading to a dominance of the higher-order modes in $\xi$ over the more regular ones. This can be understood as follows. Bounds on the r.m.s.~perturbation amplitude or r.m.s.~perturbation angle do not rule out the presence of a spike of infinitesimal thickness and unbounded height on the surface, that would lead to a diverging of $U^2$ (the regular perturbation expansion of the propulsion velocity would be invalid).

In order to guarantee  well-posedness, and consistency with the  perturbation approach,  we choose a norm based on the r.m.s.~curvature perturbation $\tilde\kappa=\kappa+1$, where $\kappa$ is the local curvature  defined as
 \begin{equation}
\pard{\mathbf{t}}{s}=\kappa\nb,
\end{equation}
with $\mathbf{t}$ the tangential unit vector in the ($\eb_r,\eb_\theta$)-plane and $s$ the curvilinear coordinate along the surface. At leading order in $\e$, the curvature perturbation is obtained as 
\begin{equation}
\tilde\kappa=\e[(1-\mu^2)\xi''-\mu\xi'+\xi],
\end{equation}
and the corresponding $\mathbf{H}$ is obtained as
\begin{subeqnarray}\label{eq:Hcurv}
H_{mn}&=&\int_{-1}^1h_{n}(\mu)h_m(\mu)\dd\mu\\
h_{j}&=&\mu L_j'+[1-j(j+1)]L_n,\quad\forall j\nonumber.
\end{subeqnarray}
$H_{mn}$ is now a full matrix whose coefficients can be computed analytically (see Appendix \ref{secB}).

Note that because the propulsion velocity is a quadratic form of the shape perturbation, the transformation $\xi\to-\xi$ leaves the propulsion velocity unchanged. Each optimal swimming velocity therefore corresponds to two different shapes obtained for $\e$ and $-\e$. Also, the change $\xi_n\to(-1)^n\xi_n$ simply performs a symmetry of the particle shape, changing the sign of the swimming velocity but not its magnitude.  

%%%
\subsubsection{Optimal swimming shape for fixed flux ($\Da=0$)}
\begin{figure}
\begin{center}
\includegraphics[width=0.4\textwidth]{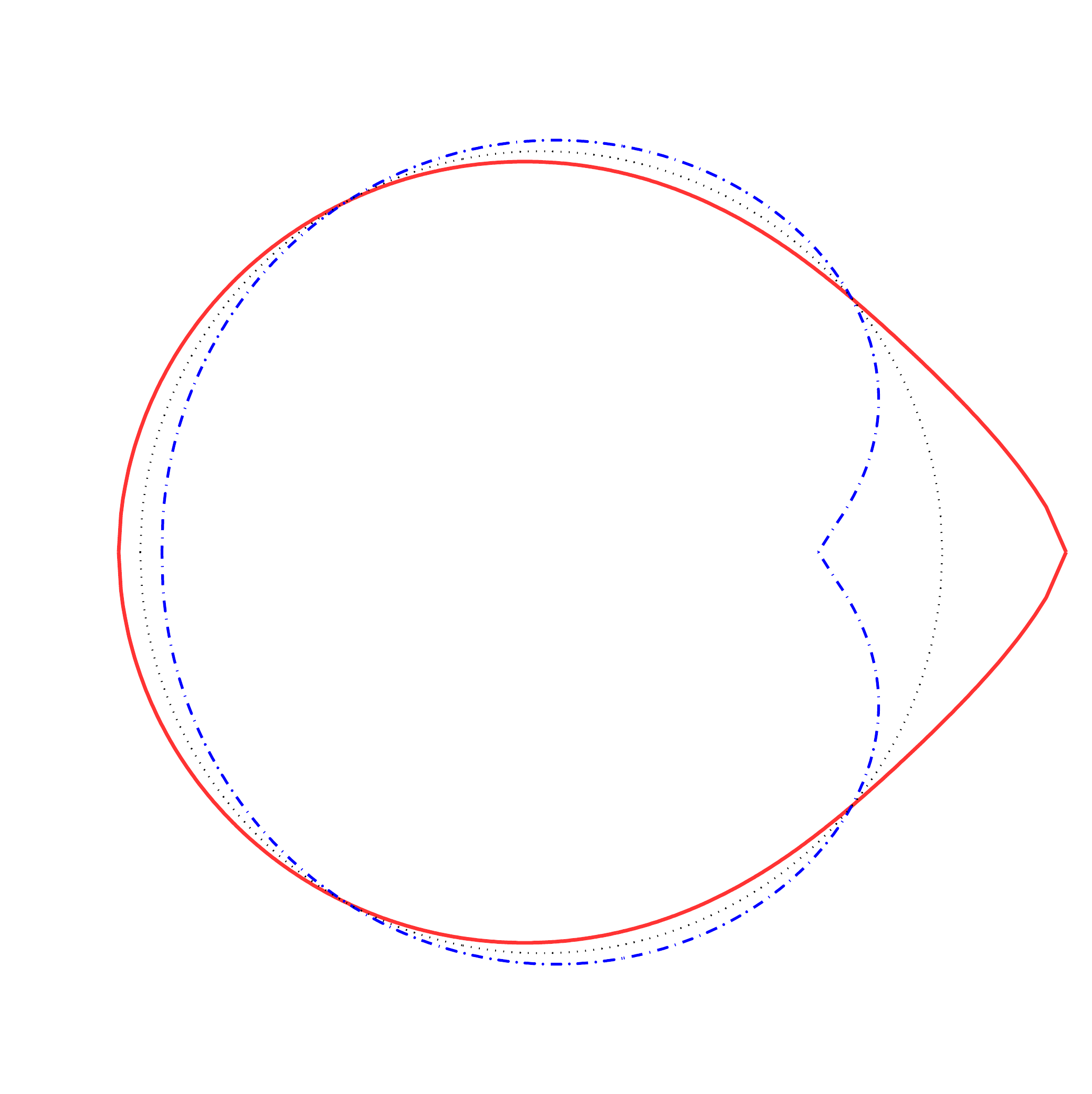} 
\caption{Small-amplitude perturbations of a sphere leading to maximum swimming velocity ($\e=0.3$ was chosen for plotting purposes). There are two solutions, one convex (solid line) and one concave (dash-dotted line), leading to the same propulsion velocity $U\approx 0.01 \e^2 M$.  The unit sphere is shown for reference (dotted line). }\label{fig:optimal}
\end{center}
\end{figure}

We plot Fig.~\ref{fig:optimal}  the two optimal shapes leading to maximum propulsion velocity at $\Da=0$ (fixed-flux absorption). For positive mobility ($M=1$), both shapes swim to the right ($U>0$) with a  small speed,  $U\approx 0.01 \e^2$. Both shapes are characterized by a sharp corner with a finite angle $\gamma$ at the pole, defined in Eq.~\eqref{eq:gammadef}. This angle is proportional to $\e$, a consequence of effectively imposing the curvature change to be $O(\e)$.

\begin{figure}
\begin{center}
\includegraphics[height=5.5cm]{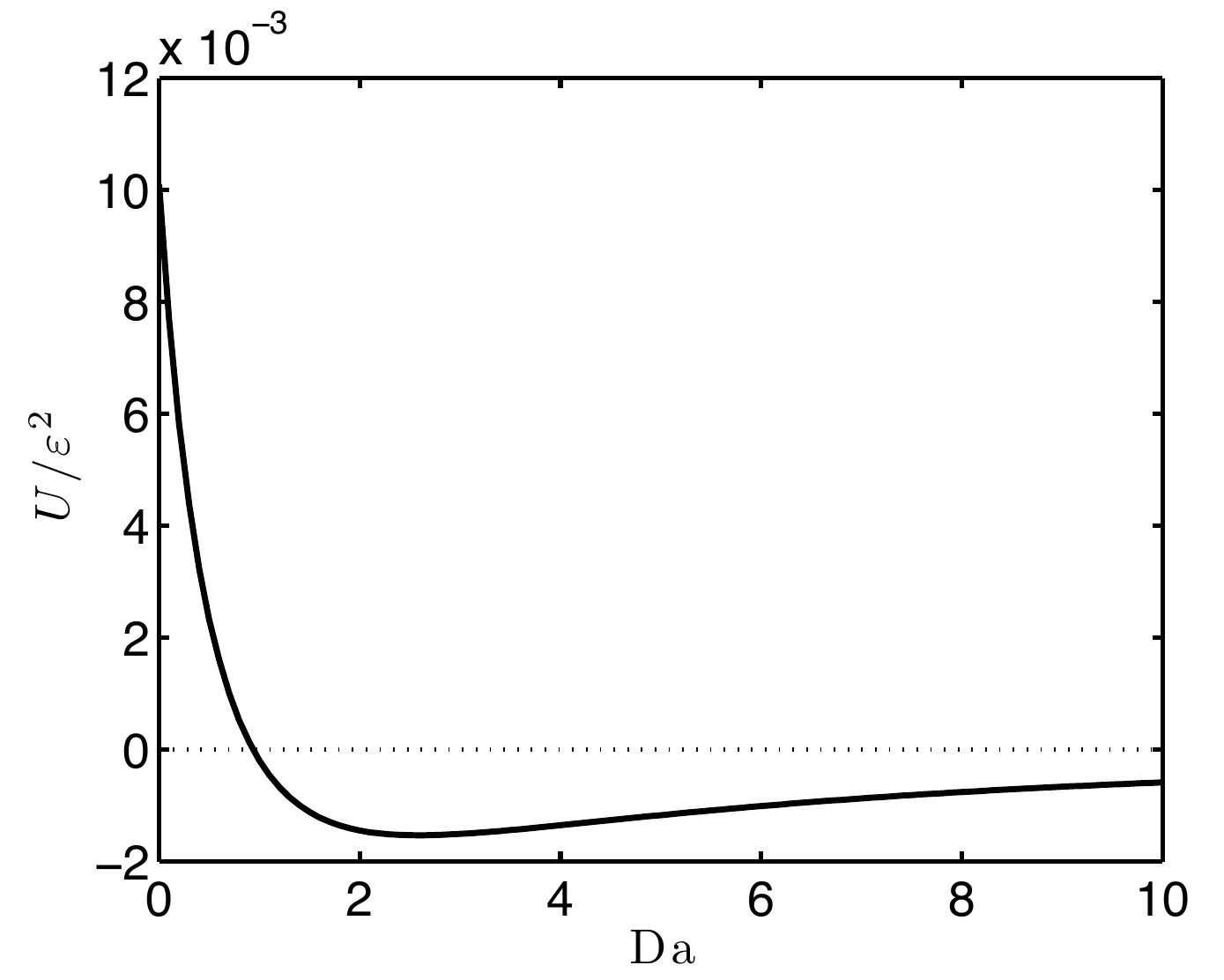}
\caption{Dependence of the swimming velocity on  $\Da$  for the optimal shapes  shown in Fig.~\ref{fig:optimal} (i.e.~optimal shape at $\Da=0$)}\label{fig:velocity_optim1_Da}
\end{center}
\end{figure}

%Near a sharp-corner generating a fixed  flux of solute, the presence of this protrusion (or intrusion) allows for a greater variation of the concentration along the surface. Typically, the concentration at the tip (resp. in the trough) is higher (resp. lower) than around the mid-section as the distance to the particle's center is higher (resp. lower). When $M>0$, this generates in turn a slip velocity directed away from the tip (resp. toward the trough). In both cases, the slip velocity is dominantly contributing from right to left, therefore generating a swimming velocity to the right. However, this attempt at understanding the direction of propulsion is flawed: the reasoning above in fact would lead to a swimming velocity at $O(\e)$ as it neglects any correction to the solute concentration (i.e. only include $c^0$) and is equivalent to neglecting all terms in Eq.~\eqref{eq:bc2_1} but the third one (in $\partial^2 c^0/\partial r\partial \mu$). The analysis of the previous sections show that this effect is cancelled out, and that many different corrections need to be taken into account to fully understand the sign and intensity of the swimming velocity. In particular, the solute concentration distribution is modified by the local fluctuation in the solid's curvature: this edge effect is a common feature of solutions of Laplace problems \citep{jackson1962}, and results in gradient enhancements near tips or wedges in the particle shape.

The optimal shapes appear to be cusped. It is a well known fact that, for example, in electrostatics, cusps can lead to divergence of electric fields, a result true in general to solutions of Laplace equation near tips or wedges in the geometry  \citep{jackson1962}. We believe that the same effect is observed in our simulations where the maximum velocity, resulting from surface gradients of the solution to Laplace's equation,  corresponds to a kinked geometry. \change{As for the two-sphere system, we note that the framework used here is valid provide that the  typical  radius of curvature near the regularized cusp  remains compared to the thickness  of the interaction layer.}

In Fig.~\ref{fig:velocity_optim1_Da} we show the dependence  of the swimming velocity associated with this optimal shape on the value of $\Da$. In general, reactive effects are observed to reduce the swimming velocity,  a property that was also observed for spherical Janus particles \citep{michelin2014} and remain valid here: reactive effects tend to reduce concentration contrasts as the chemical reaction is slowed down in regions already strongly impacted by the depletion in solute resulting from the reaction. More quantitatively, at large $\Da$, one observes that for a fixed particle shape $U\sim \Da^{-2}$. However, this decrease is not monotonous: in fact, a reversal is observed in the swimming direction as already predicted in the previous section. As a consequence, for a finite value of $\Da$ ($\Da\approx 0.94$ in this particular case), the  asymmetry-driven self-propulsion vanishes.

\subsubsection{Optimal swimming shape for arbitrary $\Da$}
\begin{figure}
\begin{center}
\includegraphics[width=.5\textwidth]{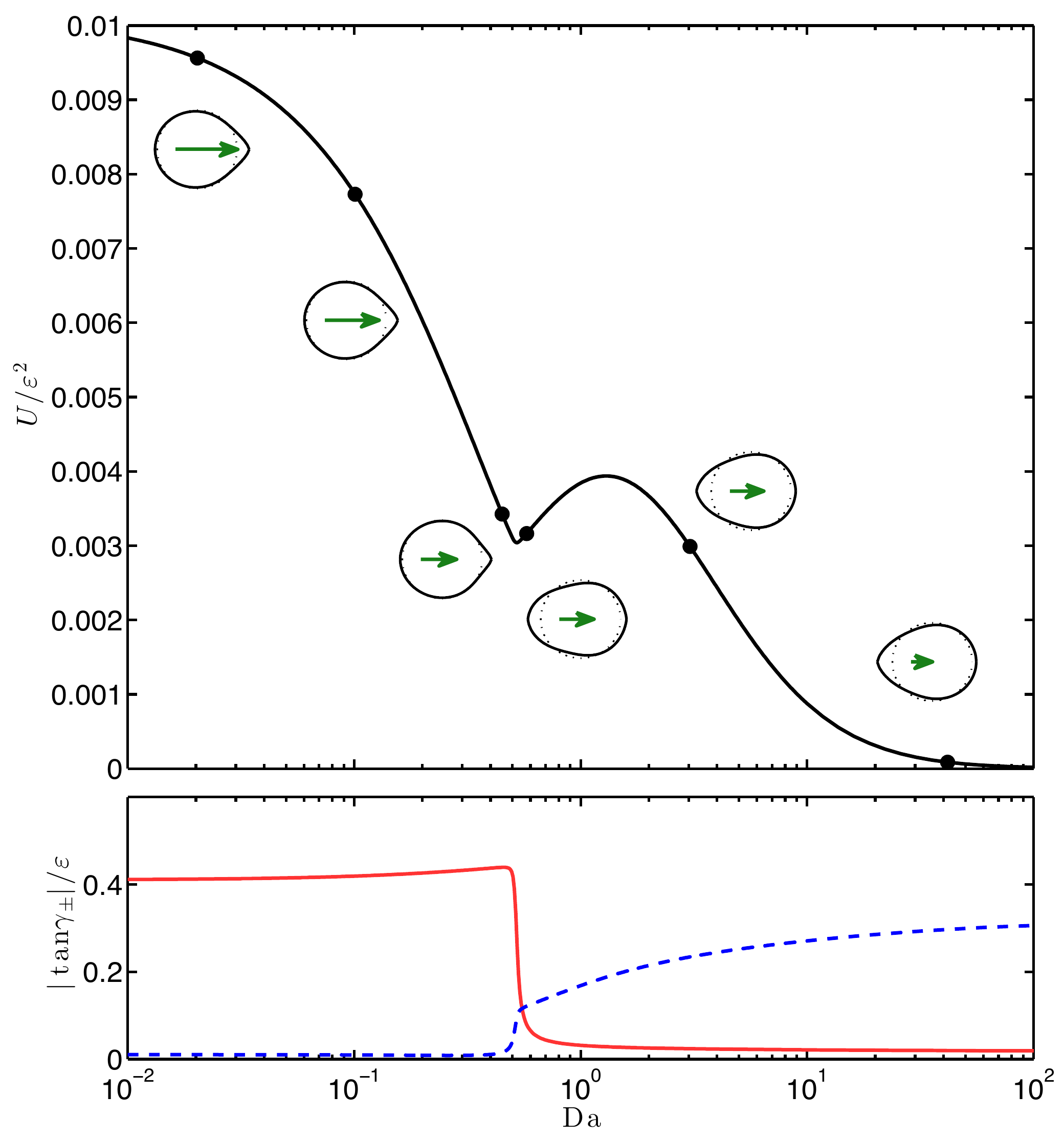}
\caption{Top: Dependence  of the optimal swimming velocity and optimal shape with $\Da$. A front-back reversal in the optimal shapes  occurs near $\Da\approx 0.5$. The direction of the swimming velocity is indicated for each displayed shape by a green arrow. Bottom: Variation  with $\Da$ of the summit angles on the optimal shape. $\gamma_+$ (solid) refers to the right pole and $\gamma_-$ (dash-dotted) refers to the left pole.}\label{fig:optimal_shapeDa}
\end{center}
\end{figure}
% Seb - si on rajoutait des flches sur les figures pour bien indiquer que la forme change de symŽtrie mais la direction de la nage reste la meme? (pour une valeur de M, disons M=1)

The previous result clearly suggests that the optimal swimming shape can not be independent of $\Da$ since any shape will perform poorly for some range of $\Da=O(1)$. Our optimization analysis can hence be extended to finite  values of the Damk\"ohler number and the results are presented on Fig.~\ref{fig:optimal_shapeDa}, where we plot the overall shape (top) and the angles at both poles (bottom). For large $\Da$, one recovers the general decay of self-propulsion velocities due to reactive effects and $U_\textrm{opt}\sim \Da^{-2}$. The optimal shape is rather insensitive to fluctuations in $\Da$ when $\Da\leq 0.4$ and $\Da\geq 0.7$. Within a narrow range of $\Da$, a sharp transition takes place. While  at lower $\Da$, the optimal shape presents a protrusion located at its front,  for greater $\Da$ the protrusion is in its back. This change in the optimal configuration is related to the change in self-propulsion direction for a particle shape with a protrusion observed in Fig.~\ref{fig:velocity_optim1_Da}. For small $\Da$, it is advantageous to have a protrusion in the front, while at larger $\Da$, the protrusion should be located in the back of the particle.

%\begin{figure}
%\begin{center}
%\includegraphics[width=.5\textwidth]{optimal_shape_angles.pdf}
%\caption{Variation  with $\Da$ of the summit angles on the optimal shape. $\gamma_+$ (solid) refers to the right pole and $\gamma_-$ (dash-dotted) refers to the left pole.}\label{fig:angle}
%\end{center}
%\end{figure}

% HERE

%%%%%%%%%%%%%%%%%%%%%%
\section{Conclusion}

The results presented in this paper for two idealized setups demonstrate that geometric asymmetries are sufficient to enable self-propulsion of a chemically-homogeneous system. As in the work of Ref.~\cite{shklyaev2014},  this identifies an alternative route to self-propulsion without exploiting chemical patterning of the particle which might otherwise be practically difficult to achieve and control. We note  however that the propulsion velocities are in general smaller than what would be achieved for an equivalent Janus system exploiting chemical asymmetry. 

The origin of this geometric self-propulsion mechanism stems from a combination of   enhanced  concentration gradients and  change in the  particle surface exposed to these  gradients. The former corresponds to an enhancement of the concentration level and gradients, and hence slip velocities, near a protrusion from a surface of constant curvature, and is a generic feature of computing local solutions to laplacian on wedge or tip singularities \citep{jackson1962}. The latter can be best illustrated in the near-sphere case: the extension of the solid particle away from its mean radius, exposes its surface to a larger range of the concentration distribution created by the particle, which modifies surface velocities. The  dependence of the final results for the swimming velocity on the various surface modes, on the deformation amplitude, and on the relative importance of reactive  effects show however that it is a nontrivial  nonlinear interplay between enhanced gradients and shape changes which results in swimming.

 Most of our results were obtained with a minimal kinetic model for the reaction at the surface of the particle, namely a fixed-flux release or absorption, but we also showed that introducing a more complex one-step reaction kinetics does not fundamentally change the results, in particular the main idea of self-propulsion using shape asymmetries.  Our work emphasizes however that the precise kinetics and the relative importance of reactive and diffusive effects may significantly influence the sign of the propulsion velocity, and additional  work will be needed on more complex, and realistic, surface chemical conditions \change{(see for example Ref.~\cite{ebbens2012})}. Furthermore, while we focused on the idea of self-propulsion in this work, the same idea will be useful as a method to apply forces on stationary bodies, leading therefore to potentially new microfluidic applications   in  pumping and mixing \cite{squires05}.

\section*{Acknowledgements}
This research was funded in part by the European Union through a Marie Curie grant to EL and by the French Ministry of Defense through a DGA (D\'el\'elation G\'en\'erale pour l'Armement) to SM).
%Seb: rajoute ton numŽro de grant

%%%%%%%%%%%%%%%%%%%%%%
\appendix

\section{Useful properties of the Legendre polynomials}
We make use of the following in the main text of the paper
\label{sec:app_formulas}
\begin{align}
&\frac{2p+1}{2}\int_{-1}^1L_nL_p\dd\mu=\delta_{np},\label{eq:leg1}\\
&\frac{2p+1}{2p(p+1)}\int_{-1}^1(1-\mu^2)L_n'L_p'\dd\mu=\delta_{np},\label{eq:leg6}\\
&\frac{2p+1}{2}\int_{-1}^1\mu L_nL_p\dd \mu=\frac{p+1}{2p+3}\delta_{n,p+1}+\frac{p}{2p-1}\delta_{n,p-1},\label{eq:leg2}\\
&\frac{2p+1}{2}\int_{-1}^1(1-\mu^2)L_n'L_p\dd\mu=\frac{(p+2)(p+1)}{2p+3}\delta_{n,p+1}\nonumber\\
&\hspace{4.5cm}-\frac{p(p-1)}{2p-1}\delta_{n,p-1},\label{eq:leg4}\\
&\frac{2p+1}{2p(p+1)}\int_{-1}^1\mu(1-\mu^2)L_n'L_p'\dd\mu=\frac{p+2}{2p+3}\delta_{n,p+1}\nonumber\\
&\hspace{5cm}+\frac{p-1}{2p-1}\delta_{n,p-1},\label{eq:leg7}\\
&\frac{2p+1}{2}\int_{-1}^1\frac{L_p\dd\mu}{\sqrt{\cosh\tau_\pm-\mu}}=\sqrt{2}\ee^{-(p+1/2)|\tau_\pm|}\label{eq:leg3}\\
&\frac{2p+1}{2}\int_{-1}^1\frac{\mu L_p\dd\mu}{\sqrt{\cosh\tau_\pm-\mu}}=\sqrt{2}\left(\frac{p+1}{2p+3}\ee^{-(p+3/2)|\tau_\pm|}\right.\nonumber\\
&\left.\hspace{4cm}+\frac{p}{2p-1}\ee^{-(p-1/2)|\tau_\pm|}\right),\label{eq:leg5}
\end{align}
\begin{align}
&\frac{2p+1}{2p(p+1)}\int_{-1}^1\frac{(1-\mu^2)L_p'\dd\mu}{\sqrt{\cosh\tau_\pm-\mu}}=\sqrt{2}\left(\frac{\ee^{-(p-1/2)|\tau_\pm|}}{2p-1}\right.\nonumber\\
&\hspace{4.5cm}\left.-\frac{\ee^{-(p+3/2)|\tau_\pm|}}{2p+3}\right).\label{eq:leg8}
\end{align}

\section{Computing the second-order swimming velocity}
\label{secB}
Applying the reciprocal theorem, Eq.~\eqref{eq:recthm}, to the second order surface velocity, Eq.~\eqref{eq:u2bc}, one obtains
\begin{align}
\Ub=&\frac{M\eb_z}{2(1+\Da)}\sum_{m,n=0}^\infty\xi_m\xi_n\left[\frac{n(n-1)(n+1)}{n+1+\Da}I_{mn}^1\right.\nonumber\\
&-\left(\frac{n-1}{n+1+\Da}\right)I_{mn}^2 -\frac{(n-1)(2n-1+\Da)}{n+1+\Da}I_{mn}^3\nonumber\\
&\left.-\frac{(n+1)(2+\Da)}{n+1+\Da}I^4_{mn}+\sum_{p=0}^\infty \frac{E_{mnp}}{p+1+\Da}I_p^5\right].\label{eq:app1}
\end{align}
where $E_{mnp}$ was defined in Eq.~\eqref{eq:emnp} and $I_{mn}^1$, $I_{mn}^2$, $I_{mn}^3$, $I_{mn}^4$ and $I^5_p$ are integrals of the Legendre polynomials that can be computed using classical properties of such polynomials and those listed in appendix~\ref{sec:app_formulas}
\begin{align}
I_{mn}^1&=\int_{-1}^1\mu L_m(\mu)L_n(\mu)\dd \mu,\\
I_{mn}^2&=\int_{-1}^1(1-\mu^2)\mu L'_m(\mu)L'_n(\mu)\dd \mu,\\
I_{mn}^3&=\int_{-1}^1(1-\mu^2)L_m(\mu)L_n'(\mu)\dd \mu,\\
I_{mn}^4&=\int_{-1}^1(1-\mu^2)L_m'(\mu)L_n(\mu)\dd \mu,\\
I_p^5&=\int_{-1}^1(1-\mu^2)L_p'(\mu)\dd \mu.
\end{align}

Also, using the definition of $E_{mn1}$ and the properties of Legendre polynomials, we obtain
\begin{equation}
E_{mn1}=\frac{3(f_n\delta_{m,n+1}+g_n\delta_{m,n-1})}{2(2m+1)(2n+1)(n+1+\Da)}
\end{equation}
with
\begin{align}
f_n=&(n+1)\Big(2n\Da^2+2n(n+4)\Da+2(n+1)(2n+1)\Big)\nonumber\\
&+n(n+1)(n+2)(n-3-\Da),\\
g_n=&n\Big(2n\Da^2+2n(n+4)\Da+2(n+1)(2n+1)\Big)\nonumber\\
&+n(n-1)(n+1)(n-3-\Da)
\end{align}

%\Big[&\Big(2n\Da^2+2n(n+4)\Da+2(n+1)(2n+1)\Big)\nonumber\\
%&\Big((n+1)\delta_{m,n+1}+n\delta_{m,n-1}\Big)\nonumber\\%
%&+n(n+1)(n-3-\Da)\Big((n+2)\delta_{m,n+1}+(n-1)\delta_{m,n-1}\Big)\Big]
%\end{align}
Substituting these results into Eq.~\eqref{eq:app1}, we obtain $U^2$ in the form
\begin{align}
U^2=&\frac{M}{(1+\Da)(2+\Da)}\sum_{m,n=0}^\infty\xi_m\xi_n\times\nonumber\\
&\hspace{2cm}\frac{\left(q_n\delta_{m,n+1}+r_n\delta_{m,n-1}\right)}{(n+1+\Da)(2n+1)(2m+1)},
\end{align}
with
\begin{align}
q_n&=-(n+1)\Big[2(n+1)\Da^2+(-2n^3+5n^2+6n+8)\Da\nonumber\\
&\hspace{1cm}-5n^3+9n^2+8n+6\Big]\\
r_n&=n\Big[2n\Da^2-(2n+3)(n^2-4n+1)\Da\nonumber\\
&\hspace{1cm}-3(n+1)(n^2-4n+1)\Big].
\end{align}
Finally,
\begin{align}
%U^2&=\frac{M}{(1+\Da)(2+\Da)}\sum_{n=0}^\infty\frac{\xi_n\xi_{n+1}}{(n+1+\Da)(n+2+\Da)(2n+1)(2n+3)}\left((n+2+\Da)q_n+(n+1+\Da)r_{n+1}\right)\nonumber\\
U^2&=\sum_{n=0}^\infty a_n(\Da)\xi_n\xi_{n+1},
\end{align}
with $a_n(\Da)$ defined in Eq.~\eqref{eq:U2}.

\section{Curvature norm tensor}\label{secC}
Choosing the r.m.s.~curvature perturbation to define the $\mathbf{H}$ operator in Eq.~\eqref{eq:evp}, the symmetric tensor $H_{mn}$ is defined in Eq.~\eqref{eq:Hcurv} and is symmetric. In particular, using classical properties of Legendre polynomials (see Appendix~\ref{sec:app_formulas}),  it can be computed exactly , when $m\leq n$, as
\begin{align}
H_{mn}=&\left[\frac{n^4-4n^2-n+1}{2n+1}+2b_n\right]\delta_{mn}\nonumber\\
&+2\sum_{q=0}^{2q\leq n-2}\left(1-m(m+1)+b_m\right)\delta_{m,n-2q-2},
\end{align}
with
\begin{equation}
b_n=2\sum_{p=0}^{2p\leq n-1}(2n-4p-1).
\end{equation}

%\newpage

\end{document}